\documentclass[linenumbers,referee]{raa}
\usepackage[utf8]{inputenc}
\usepackage{graphics,graphicx}
\usepackage{url}
\usepackage{wrapfig}
\usepackage{float}
\usepackage{booktabs}
\usepackage{threeparttable}
\usepackage{refcount}
\usepackage{authblk}
\usepackage{orcidlink}
\usepackage{natbib}
\usepackage{amsmath}
\usepackage{amssymb}
\usepackage{tabularx}
\newcolumntype{C}{>{\centering\arraybackslash}X} 
\usepackage{hyperref}

\begin{document}

\titlerunning{Euclid/CSST/JPCam/JUST cosmic noon survey}
\authorrunning{J.-T. Li et al.}

\title{Probing Large-scale Structure and the Multi-Phase IGM at the Cosmic Noon \\
\large --- Insights from a Joint Survey with Euclid, CSST, JPCam, and JUST}

\author{Jiang-Tao Li\inst{1}\orcidlink{0000-0001-6239-3821}\thanks{Corresponding author: pandataotao@gmail.com}
\and
Renato A. Dupke\inst{2,3}
\and
Yan Gong\inst{4,5}\orcidlink{0000-0003-0709-0101}
\and
Zhijie Qu\inst{6}\orcidlink{0000-0002-2941-646X}
\and
Weichen Wang\inst{7}\orcidlink{0000-0002-9593-8274}
\and 
Xiaohu Yang\inst{8,9}\orcidlink{0000-0003-3997-4606}
\and
Xiaodi Yu\inst{10}\orcidlink{0009-0004-1889-3043}
}

\institute{Purple Mountain Observatory, Chinese Academy of Sciences, 10 Yuanhua Road, Nanjing 210023, People’s Republic of China
\and
Observat\'{o}rio Nacional, Rua Gal.\ Jos\'{e} Cristino 77, S\~{a}o Crist\'{o}v\~{a}o, 20921--400 Rio de Janeiro, Brazil
\and
University of Michigan, Department of Astronomy, 1085 South University Ave., Ann Arbor, MI, United States of America
\and
National Astronomical Observatories, Chinese Academy of Sciences, 20A Datun Road, Beijing 100101, People’s Republic of China
\and
School of Astronomy and Space Science, University of Chinese Academy of Sciences(UCAS), Yuquan Road No.19A Beijing 100049, People’s Republic of China
\and
Department of Astronomy, Tsinghua University, Beĳing 100084, People’s Republic of China
\and
Department of Physics, University of Milano-Bicocca, Piazza della Scienza 3, I-20126 Milano, Italy
\and
State Key Laboratory of Dark Matter Physics, Tsung-Dao Lee Institute \& School of Physics and Astronomy, Shanghai Jiao Tong University, Shanghai 201210, People’s Republic of China
\and
Shanghai Key Laboratory for Particle Physics and Cosmology, and Key Laboratory for Particle Physics, Astrophysics and Cosmology, Ministry of Education,  Shanghai Jiao Tong University, Shanghai 200240, People’s Republic of China
\and
Institute of Astronomy and Information, Dali University, Dali 671003, People’s Republic of China
}

\abstract{We present scientific and technical justifications of a potential coordinated Euclid/CSST/JPCam/JUST survey of the Euclid Deep Field North (EDF-N), aimed at probing the multi-phase circumgalactic and intergalactic medium (CGM/IGM) at the cosmic noon over $\sim20\rm~deg^2$. The survey is structured around three connected goals: (1) improving photometric redshift (photo-$z$) accuracy through the combination of broad- and narrow-band photometry, enabling reliable identification of large-scale structures; (2) probing extended CGM emission with dedicated narrow-band imaging; and (3) mapping foreground IGM via absorption-line spectroscopy of background galaxies. Together, these components establish an integrated observational framework to investigate galactic ecosystems --- linking galaxies to their circumgalactic and intergalactic environments --- at cosmic noon. We show that the J-PAS-like narrow-band system used in JPCam substantially improves photo-$z$ accuracies from only the Euclid/CSST broad-band data, especially for star-forming galaxies at $z\sim1.0-1.4$. This enables the identification of galaxy groups and (proto-)clusters directly from photo-$z$ measurements. Stacked JPCam narrow-band imaging should also detect extended [\ion{O}{II}]-emitting CGM halos. We then construct mock three-dimensional (3D) gas distribution model and realistic galaxy catalog, and further construct mock CSST and JUST background galaxy spectra adding Ly$\alpha$ and \ion{Mg}{II} absorptions. The reconstructed 3D \ion{H}{I} field from CSST Ly$\alpha$ forest reliably recovers large-scale structures; however, our simulations indicate that detecting diffuse IGM \ion{Mg}{II} absorption with JUST is infeasible, either through spectral stacking or via the two-point correlation function method. We conclude that constraining the metallicity of the diffuse IGM will require significantly deeper and higher-resolution spectroscopy expected from future facilities such as the 39m E-ELT.
}

\maketitle

\section{Introduction} \label{sec:Introduction}

Galaxies are embedded within a complex ``galactic ecosystems'' --- a multi-scale network composed of both gaseous and structural components. The gaseous elements include the circumgalactic medium (CGM) and the intergalactic medium (IGM), while the structural elements encompass galaxies, active galactic nuclei (AGN), galaxy groups, and clusters. All of these components are interconnected and organized by the underlying cosmic web, whose large-scale matter distribution shapes the flow of gas and the growth of galaxies. This ecosystems governs galaxy formation and evolution across cosmic time. In this work, we use the term ``galactic ecosystems'' to emphasize the coupled evolution of galaxies and their surrounding gaseous environments across multiple spatial scales, from the CGM to the large-scale IGM. The epoch of cosmic noon ($z\sim1-2$) is particularly important, marking the peak of global star formation (SF) activity (e.g., \citealt{Dickinson2003, Madau2014}). Understanding the exchange of matter, metals, and energy between galaxies and their surrounding gaseous reservoirs during this era is therefore essential for explaining how such vigorous SF was both sustained and regulated. The observational strategy developed in this paper is specifically designed to probe these interconnected components in a unified manner.

In principle, the CGM/IGM over a broad temperature range (e.g., $T<10^6\rm~K$) can be probed through both absorption and emission lines in the rest-frame ultraviolet (UV)/optical bands. Absorption-line studies typically reach lower gas column densities, while emission-line mapping provides improved spatial sampling of denser structures. In particular, the IGM tomography technique has emerged as an effective method to probe large-scale diffuse gas across a broad redshift range including the cosmic noon, using background quasars or galaxies as light sources to detect rest-frame UV absorption lines mainly from the IGM (e.g., \citealt{Lee2014, Pieri2016, Japelj2019}). The Ly$\alpha$ forest provides a powerful means of reconstructing the three-dimensional (3D) cosmic web (e.g., \citealt{Lee2018}), while metal lines such as \ion{C}{IV}, \ion{Si}{IV}, and \ion{Mg}{II} trace chemical enrichment and feedback driven outflows, offering key insights into the physical and chemical states of the CGM and IGM (e.g., \citealt{LiJ2025, Yu2025, Mao2025}). On smaller scales, closely separated sightlines or stacked galaxy-quasar pairs enable detailed characterization of the CGM and intragroup/intracluster gas (e.g., \citealt{Steidel2010, Pieri2014, Chen2020}).

A complementary avenue is to observe the gas directly in emission. Deep narrow-band imaging has proven remarkably effective for detecting extended nebulae at $z\sim1-2$ through tracers such as Ly$\alpha$, [\ion{O}{II}], [\ion{O}{III}], and \ion{He}{II} (e.g., \citealt{Cantalupo2014, Cantalupo2019, Cai2017}). Follow-up integral field spectroscopy can then map kinematics and chemical composition, revealing how SF and AGN activity shape the CGM and IGM (e.g., \citealt{Bacon2015, Borisova2016, ArrigoniBattaia2019}).

Galaxy clusters --- the most massive gravitationally bound structures in the Universe --- and their progenitors, proto-clusters, serve as key laboratories for studying the connection between galaxies and large-scale structures. Some proto-clusters likely began assembling as early as in the \ion{H}{I} reionization epoch (e.g., \citealt{Hu2021, WangF2024a}), with their intracluster medium (ICM) gradually heated through gas accretion, gravitational collapse, and feedback. The earliest systems with detectable hot, X-ray-emitting ICM likely emerged by $z\gtrsim2$ (e.g., \citealt{WangT2016, Tozzi2022, Overzier2016}). Whether fully virialized or still in the process of assembly, these overdensities strongly influence galaxy evolution through ram-pressure stripping, strangulation, tidal and galaxy-galaxy interactions, and other environmental mechanisms (e.g., \citealt{Kravtsov2012, Chiang2017}). Reliable identification of such overdense structures at $z\sim1$--2 critically depends on accurate photo-$z$ with sufficient precision to resolve structures along the line of sight. Combining IGM tomography, metal-line absorption, and emission-line mapping within these dense environments provides a uniquely comprehensive view of how galaxies co-evolve with, and are regulated by, their surrounding large-scale structures.


A new generation of wide-field photometric surveys is poised to revolutionize studies of AGN, galaxies, and large-scale structure at the cosmic noon (Table~\ref{tab:photometrysurveys}). The Vera C. Rubin Observatory’s Legacy Survey of Space and Time (LSST; \citealt{Ivezic2019}) will obtain deep, multi-epoch optical imaging over $\sim18,000\rm~deg^{2}$, far exceeding the depth and cadence achieved by SDSS \citep{York2000} and Pan-STARRS \citep{Chambers2016}. Complementary ground-based facilities such as the Wide-Field Survey Telescope (WFST; \citealt{Wang2023}) and the Chinese Space Station Survey Telescope (CSST; \citealt{Zhan2021,CSST2025}) will deliver deep, high-cadence or high-resolution optical imaging across the northern sky. Space-based near-infrared (near-IR) missions including Euclid \citep{Laureijs2011} and the Nancy Grace Roman Space Telescope \citep{Spergel2015} will provide wide-area coverage with excellent photometric-redshift (photo-$z$) performance. In addition, the Javalambre Physics of the Accelerating Universe Astrophysical Survey (J-PAS; \citealt{Benitez2014}) using the Javalambre Survey Telescope (JST250) and the JPCam camera will obtain low-resolution ``spectral'' imaging using 56 contiguous narrow bands across thousands of square degrees, while the Dark Energy Survey (DES; \citealt{DES2021}) continues to supply deep multi-band optical imaging over $\sim5000\rm~deg^{2}$. Together, these surveys will identify unprecedented samples of AGN and galaxies over a broad redshift range, deliver photo-$z$ of sufficient accuracy to trace large-scale structure, and map the cosmic web across Gpc scales. Compared to earlier surveys such as SDSS and Pan-STARRS, they offer order-of-magnitude gains in depth and photometric quality over comparable or larger sky areas, enabling transformative investigations of galaxies and large-scale structures during the epoch of peak cosmic SF. These datasets typically achieve $\sigma_z/(1+z)\sim0.01$--0.05 for galaxies at $z\sim1$--2, depending on filter configuration and depth, which is adequate for statistical large-scale structure studies but still limits the identification of lower-mass groups and proto-clusters.

\begin{table*}[htbp]
\centering
\begin{threeparttable}
\caption{Comparison of major wide-field photometric surveys relevant for studies of galaxies, AGN, and large-scale structure at $z\sim1$--2.}
\label{tab:photometrysurveys}
\begin{tabularx}{\textwidth}{lCCCCCCC}
\hline
\textbf{Survey$^{\rm a}$} &
\textbf{Start} &
\textbf{Area} [deg$^2$] &
\textbf{Depth$^{\rm b}$} [mag] &
\textbf{Bands} &
\textbf{$\lambda$} [nm] &
\textbf{Resolution} [$^{\prime\prime}$] &
\textbf{Cadence} [days] \\
\hline
SDSS       & 2000 & $\sim$14{,}000 & $r\simeq22.2$            & $u,g,r,i,z$               & 300--1000 & $\sim$1.4  & -- \\
Pan-STARRS & 2010 & $\sim$30{,}000 & $r\simeq23.2$            & $g,r,i,z,y$               & 400--1000 & $\sim$1.1  & $\sim3$$^{\rm d}$ \\
DES        & 2013 & $\sim$5{,}000  & $r\simeq24.1$            & $g,r,i,z,Y$               & 400--1000 & $\sim$0.96 & $\sim7$ \\
LSST       & 2025 & $\sim$18{,}000 & $r\simeq27.5$            & $u,g,r,i,z,y$             & 320--1050 & $\sim$0.7  & $\sim3$ \\
WFST       & 2023 & $\sim$6{,}000 & $r\simeq22.8$             & $u,g,r,i$                 & 320--1000 & $\sim$1.0  & $\sim10$ \\
CSST       & 2027 & $\sim$17{,}500 & $r\simeq26.0$            & NUV, $u,g,r,i,z,y$        & 255--1065 & $\sim$0.15 & -- \\
Euclid     & 2024 & $\sim$14{,}000 & ${\rm VIS}\simeq24.5$    & VIS + $Y,J,H$             & 550--2000 & $\sim$0.18 & -- \\
Roman      & 2027 & $\sim$2{,}400  & $\sim$26.5$^{\rm c}$     & 8 filters                 & 480--2300 & $\sim$0.11 & -- \\
J-PAS      & 2023 & $\sim$8{,}500  & $r\simeq23.5$            & 54 NB + MB/BB             & 350--910  & $\sim$1.0  & -- \\
\hline
\end{tabularx}

\begin{tablenotes}[flushleft]\footnotesize
\item[a] The (expected) start year, area, depth, and cadence values listed here are characteristic values adopted for general reference. Many surveys have multiple observing modes or tiers with different footprints, depths, and revisit strategies; the numbers quoted here correspond to commonly referenced wide or main survey configurations.
\item[b] Depths are approximate $5\sigma$ point-source limits in AB magnitudes unless otherwise noted; exact values depend on the adopted S/N definition, source morphology, aperture choice, and coaddition strategy.
\item[c] For space-based missions without a standard Sloan $r$ band (e.g., Euclid and Roman), the quoted depth refers to a representative optical or near-infrared imaging depth commonly adopted in the literature (e.g., Euclid VIS or Roman High Latitude Survey imaging).
\item[d] The cadence of Pan-STARRS quoted here refers to the PS1 Medium Deep Survey (MDS) which has the highest cadence among all Pan-STARRS surveys (e.g., \citealt{Chambers2016}). It is not for the same survey as the Area and Depth columns.
\end{tablenotes}

\end{threeparttable}
\end{table*}


A parallel revolution is underway in wide-field, multi-object spectroscopic surveys (Table~\ref{tab:spectroscopicsurveys}), which will advance studies of AGN, galaxies, and large-scale structure while also enabling absorption-line measurements of diffuse gaseous environments such as the CGM, IGM, and the cosmic web. The Dark Energy Spectroscopic Instrument (DESI; \citealt{DESI2016}) is obtaining $\sim35$ million spectra over $\sim14,000\rm~deg^{2}$, yielding an unprecedented density of background quasars and luminous galaxies for Ly$\alpha$ forest and metal absorption line analyses. The Subaru Prime Focus Spectrograph (PFS; \citealt{Takada2014}) will reach significantly deeper detection limit ($r\sim24\rm~mag$) over $\sim1,400\rm~deg^{2}$ with broad optical-near-IR coverage. Large-area surveys with 4MOST \citep{deJong2019} and WEAVE \citep{Dalton2016} will obtain millions of medium- and high-resolution spectra, substantially expanding the number of background sources suitable for precision IGM tomography. Emerging Chinese facilities --- including the 4.4m Jiaotong University Spectroscopic Telescope (JUST; \citealt{JUST2024}) and the 6.5m MUltiplexed Survey Telescope (MUST; \citealt{Zhao2024,Cai2025}) --- are designed for high-multiplex, medium- to high-resolution spectroscopy, and will provide extensive samples of background quasars and star-forming galaxies, particularly across the northern sky. Collectively, these programs will supply orders-of-magnitude larger and denser grids of background sightlines than previous generations of surveys (e.g., SDSS-V; \citealt{Kollmeier2017}), enabling precision 3D mapping of the CGM and IGM during the epoch of peak cosmic SF.

\begin{table*}[htbp]
\centering
\begin{threeparttable}
\caption{Representative capabilities of major wide-field spectroscopic surveys relevant for studies of AGN, galaxies, and large-scale structures at $z\sim1-2$. Listed values are approximate design or survey goal parameters.}
\label{tab:spectroscopicsurveys}
\begin{tabularx}{\textwidth}{lCCCCCCC}
\hline
\textbf{Survey} &
\textbf{Start} &
\textbf{Area} [deg$^2$] &
\textbf{Targets} [$10^6$] &
\textbf{$\lambda$} [nm] &
\textbf{$\lambda/\Delta\lambda$} &
\textbf{Multiplex} &
\textbf{References} \\
\hline
SDSS-V   & 2020 & $\sim$14{,}000 & $\sim$6  & 360--1000 & $R\sim2000$          & $\sim$5000 & \citet{Kollmeier2017} \\
DESI     & 2021 & $\sim$14{,}000 & $\sim$35 & 360--980  & $R\sim2000$--5000    & 5000       & \citet{DESI2016} \\
PFS      & 2025 & $\sim$1{,}400  & $\sim$4--5 & 380--1260 & $R\sim2300$--5000    & $\sim$2400 & \citet{Takada2014} \\
4MOST    & 2026 & $\sim$15{,}000 & $\sim$25--30 & 370--950 & $R\sim5000$ / 20{,}000 & $\sim$2400 & \citet{deJong2019} \\
WEAVE    & 2023 & $\sim$10{,}000 & $\sim$10--15 & 370--960 & $R\sim5000$ / 20{,}000 & $\sim$1000 & \citet{Dalton2016} \\
JUST     & 2027 & $\sim$14{,}000 & $\sim$10 & 365--925 & $R\sim2000$--5000    & $\sim$2000 & \citet{JUST2024} \\
MUST     & 2031 & $\sim$5{,}000  & $\sim$10 & 350--1000 & $R\sim3000$--20{,}000 & $\sim$4000 & \citet{Cai2025} \\
\hline
\end{tabularx}

\begin{tablenotes}[flushleft]\footnotesize
\item[a] The (expected) start year, area, target numbers, and multiplex values listed here are characteristic survey-scale values intended for general comparison. Many facilities operate multiple survey programs with different footprints, depths, and target-selection strategies.
\item[b] Target numbers denote the total planned number of unique objects observed over the full survey duration; exact values depend on survey tier, exposure strategy, and target class (e.g., galaxies, AGN, stars).
\item[c] Spectral resolution ranges reflect the availability of multiple spectrograph modes (e.g., low- and high-resolution arms) within a given facility.
\item[d] For planned or proposed facilities (e.g., JUST and MUST), values represent current design goals and may evolve as instrument design and survey strategy are finalized.
\end{tablenotes}

\end{threeparttable}
\end{table*}


In this paper, we outline and justify a comprehensive strategy for studying the galactic ecosystems at cosmic noon, combining photometric observations of galaxies, AGN, and large-scale structures such as (proto-)clusters with spectroscopic and narrow-band imaging observations of extended gaseous environments. Our framework incorporates current state-of-the-art or next-generation broad-band photometric surveys from Euclid and CSST together with the narrow-band system of JST250/JPCam, whose synergy enables highly accurate photometric redshifts. We also explore the potential of narrow-band imaging to map extended emission-line structures and of follow-up spectroscopy with CSST and JUST to measure absorption lines that probe diffuse gas at lower column densities. The overall survey design is introduced in \textsection\ref{sec:survey}, followed by detailed discussions and technical justifications for the photometric and spectroscopic components in \textsection\ref{sec:PhotometricSurvey} and \textsection\ref{sec:SpectroscopicSurvey}, respectively. A summary and future prospects are presented in \textsection\ref{sec:SummaryProspects}. Unless otherwise noted, we adopt a standard $\Lambda$CDM cosmology with $(H_0, \Omega_m, \Omega_\Lambda) = (69.6~\mathrm{km~s^{-1}~Mpc^{-1}},0.286,0.714)$ (e.g., \citealt{Komatsu2011}) throughout this paper.

\section{Survey Design} \label{sec:survey}

\subsection{Overall Strategy and Field Selection} \label{subsec:Strategy}

We propose a coordinated program using Euclid, CSST, JPCam, and JUST to investigate the galactic ecosystems at cosmic noon. The science objectives are threefold:
(1) to map the environments of galaxies by identifying and characterizing large-scale structures such as (proto-)clusters and the cosmic web;
(2) to search for extended CGM emission using narrow-band imaging of key optical/UV diagnostic lines; and
(3) to characterize large-scale gaseous structures through UV absorption lines in the spectra of background sources.
This combined approach bridges a crucial redshift interval between future deep \ion{H}{I} 21-cm line surveys at lower redshifts ($z\lesssim1$; e.g., \citealt{Duffy2012}) or during the epoch of reionization ($z\gtrsim6$; e.g., \citealt{Greig2021}), and high-redshift ($z\gtrsim2$) Ly$\alpha$ forest studies that will be enabled by next-generation 30m class telescopes (e.g., \citealt{Japelj2019}).

To achieve these goals, the program combines multi-band photometry with targeted spectroscopy of bright background sources across several current and next-generation facilities. Deep imaging from Euclid, CSST, and JPCam will provide accurate photometric redshifts for AGN and galaxies, as well as estimates of stellar masses and star-formation rates (SFRs), enabling both the reconstruction of large-scale structure and the efficient selection of optimal background objects for absorption-line studies. The narrow-band system of JPCam will further enable searches for diffuse CGM emission, both around individual systems in exceptional cases and statistically through stacking analyses. Two complementary spectroscopic components are envisioned. First, low-resolution near-UV slitless spectroscopy from CSST will probe the Ly$\alpha$ forest at $z\gtrsim1.1$, and even provide a marginal baryon acoustic oscillations (BAO) detection \citep{Tan2025}. Second, because Ly$\alpha$ at these redshifts lies blueward of the optical window, ground-based spectroscopy from surveys such as JUST will target key UV metal transitions --- including \ion{Mg}{II}, \ion{C}{IV}, \ion{Si}{IV}, and others (e.g., \citealt{LiJ2025, Yu2025}) --- to trace the multi-phase IGM and CGM. To achieve fine spatial sampling of the foreground gas, the program will include all sufficiently bright background sources, most of them are not quasars but relatively massive galaxies (e.g., \citealt{Japelj2019}).

Although the choice of survey field is flexible and does not strongly constrain the scientific goals or technical requirements, we adopt the Euclid Deep Field North (EDF-N) for this study to facilitate coordinated follow-up with JUST and other survey programs. EDF-N is one of the three ultra-deep legacy fields defined by the Euclid mission, covering $\sim20\rm~deg^{2}$ near the Northern Ecliptic Pole. These deep fields are designed to anchor Euclid’s legacy science by providing imaging and near-IR slitless spectroscopy more than two magnitudes deeper than the wide survey (e.g., \citealt{Euclid2022}). Its year-round visibility enables repeated observations that enhance depth, temporal sampling, and photometric calibration. This exceptional depth and repeated coverage, combined with accessibility to complementary facilities, make EDF-N particularly well suited for coordinated CGM/IGM studies requiring accurate photometric redshifts and spectroscopic follow-up. Although EDF-N is not currently included in the J-PAS footprint, it is technically observable, and can be proposed through the OAJ Open Time program \footnote{\url{https://oajweb.cefca.es/observingtime/applying_observing_time}\label{fn:OAJOpenCall}}. It also lies within the planned CSST survey footprint \citep{Gong2019}, with final depth still to be determined. The field is also fully accessible to JUST \citep{JUST2024}, making it an excellent choice for coordinated multi-wavelength and spectroscopic follow-up, particularly with facilities in China.

\subsection{Scientific Objectives} \label{subsec:SciObj}

We summarize below the specific scientific objectives of the proposed program.

$\bullet$ \emph{Galaxies and large-scale galaxy structures.}
Characterizing galaxies and their large-scale assemblies --- galaxy groups, (proto-)clusters, and the cosmic webs --- is essential for understanding the environments that shape galaxy formation and evolution. These measurements also allow us to link the observed gaseous structures, whether on large scales in the IGM or smaller scales in the CGM, to their associated galaxy overdensities. To determine galaxy properties such as redshift, stellar mass, and SFR, we will combine typically more accurate spectroscopic data from CSST and JUST with typically deeper broad- and narrow-band photometric data from Euclid, CSST, and JPCam.

Spectroscopic observations provide the most precise redshifts but are less efficient than photometric surveys for building large statistical samples. CSST slitless spectroscopy reaches a typical detection limit of $\lesssim22\rm~mag$ \citep{Gong2019}, significantly shallower than the $\gtrsim24\rm~mag$ depths of the planned photometric surveys (e.g., \citealt{Benitez2014, Gong2019, Euclid2022}), and is therefore primarily useful for the brightest galaxies. JUST spectroscopy, while higher resolution and slightly deeper ($\sim23.5\rm~mag$; see \textsection\ref{subsec:CoupleGasGalaxy}), will be used primarily for targeted follow-up of selected background sources and a modest subset of foreground galaxies. It remains less efficient for complete galaxy structure mapping.

Therefore, deep multi-band photometry will serve as the primary channel for identifying galaxies and large-scale structures. Euclid provides the highest sensitivity in the near-IR (extending to $\sim2\rm~\mu m$), CSST adds essential blue and near-UV coverage (down to $\sim255\rm~nm$), and JPCam contributes contiguous narrow-band coverage enabling low-resolution spectral sampling (similar approaches has also been adopted in other surveys; e.g., \citealt{Ramakrishnan2025}). For example, J-PAS traces [\ion{O}{II}] and other strong emission lines across redshift intervals of $\Delta z\sim0.2$ (\textsection\ref{subsec:ExpPhotometry}), and achieves photometric redshift precision of approximately $\sigma_z\sim0.003(1+z)$ for bright enough galaxies (e.g., \citealt{Benitez2014,Bonoli2021,HernanCaballero2021}). Together, these facilities will produce accurate 3D maps of galaxies, groups, and (proto-)clusters, providing the environmental context required to interpret the CGM and IGM structures traced by absorption- and emission-line measurements.

$\bullet$ \emph{CGM.}  
The absorption lines in the spectra of bright background sources provide a powerful probe of the CGM of galaxies at the cosmic noon. Because the density of background sightlines will not be high enough to resolve the CGM of individual galaxies, we will instead identify well-defined foreground galaxy samples and statistically characterize the gas distribution by stacking galaxy-background sightline pairs to construct an effective or ``composite'' galaxy profile (e.g., \citealt{Chen2020}). This widely used technique in CGM studies combines many galaxy/AGN pairs to recover average radial profiles of gas properties. Absorption lines from ions of different ionization states will further reveal the multi-phase structure of the CGM.

In addition to absorption-line measurements, narrow-band imaging from JPCam will enable searches for diffuse CGM emission. Such emission can sometimes be detectable around individual galaxies or AGN when exceptionally bright (e.g., \citealt{Farina2017}), but will more often be recovered statistically via stacking analyses (e.g., \citealt{Matsuda2012}). Unlike most existing narrow-band studies, which rely primarily on Ly$\alpha$ at higher redshifts, our program focuses on lower redshifts and targets strong optical/UV nebular lines such as [\ion{O}{II}] $\lambda\lambda$3726.0,3728.8~\text{\AA}, \ion{C}{IV} $\lambda\lambda$1548.2,1550.8~\text{\AA}, \ion{He}{II} $\lambda$1640.4~\text{\AA}, and \ion{Mg}{II} $\lambda\lambda$2795.5,2802.7~\text{\AA} (e.g., \citealt{Stroe2017,Rupke2019,Dutta2023,Dutta2024}). The contiguous narrow-band filters of JPCam provide continuous spectral coverage, enabling efficient searches for extended emission across a wide redshift range and possibly offering new insights into the multi-phase, spatially extended CGM. Deep integral-field spectroscopy on large-aperture telescopes can achieve superior surface-brightness sensitivity for individual systems (e.g., \citealt{Rupke2019,Nielsen2024,Zhang2024}), but typically over limited fields of view and smaller samples. In contrast, wide-field narrow-band imaging discussed here enables efficient surveys of much larger galaxy samples, although with typically poorer sensitivities for individual objects, making it particularly powerful for statistical detection through stacking and for mapping CGM properties across diverse environments.

$\bullet$ \emph{IGM.}  
The 3D structure of the IGM can be mapped using IGM tomography, which exploits bright AGN and galaxies as background sources to probe intervening gas through rest-frame UV absorption lines (e.g., \citealt{Lee2014, Pieri2016, Japelj2019}). The Ly$\alpha$ forest at rest-frame $\lambda<1216~\text{\AA}$ remains the most widely used tracer of large-scale structure (e.g., \citealt{Lee2018,Newman2020}). At cosmic noon, however, Ly$\alpha$ falls in the near-UV, requiring space-based observations. In our program, Ly$\alpha$ absorption can be detected with the CSST slitless spectrograph, but only at low spectral resolution ($R=\lambda/\Delta\lambda\sim200$, corresponding to $\Delta z\sim0.01$ at $z\sim1$ or $\Delta v\sim1500\rm~km~s^{-1}$) and at relatively shallow detection limits ($\sim20.5\rm~mag$ for the wide survey and at least 1~mag deeper in the deep survey; $\sim5~\sigma$ AB mag per resolution element), and only for redshifts $z\gtrsim1.1$ \citep{Gong2019}.

An alternative way to ``map'' the IGM via absorption lines is to leverage metal-line absorption features accessible from the ground with bigger telescopes such as JUST. At $z\sim1-2$, most of the important rest-frame UV transitions fall within the optical observing window, including low-ionization tracers such as \ion{Si}{II} $\lambda\lambda$1260.4, 1304.4~\text{\AA}, \ion{C}{II} $\lambda\lambda$1334.5, 1335.7~\text{\AA}, \ion{Fe}{II} $\lambda\lambda\lambda\lambda\lambda\lambda$1608.5, 2344.2, 2374.5, 2382.8, 2586.7, 2600.2~\text{\AA}, and \ion{Mg}{II} $\lambda\lambda$2795.5, 2802.7~\text{\AA}, as well as high-ionization species such as \ion{N}{V} $\lambda\lambda$1238.8, 1242.8~\text{\AA}, \ion{Si}{IV} $\lambda\lambda$1393.8, 1402.8~\text{\AA}, and \ion{C}{IV} $\lambda\lambda$1548.2, 1550.8~\text{\AA} (e.g., \citealt{LiJ2025,Mishra2025}). These multiplets, with well-defined wavelength separations and flux ratios, provide robust diagnostics of multi-phase gas over a broad range of densities and ionization states.

Absorption-line datasets can be analyzed in three complementary ways:
(1) Statistical analyses of absorber incidence, equivalent-width distributions, and column density functions (e.g., \citealt{Yu2025});
(2) Direct tomographic reconstruction of the IGM using a dense grid of background sightlines (e.g., \citealt{Newman2020}); and
(3) Techniques for recovering weak signals from large absorber samples, such as spectral stacking (e.g., \citealt{Pieri2014}) or the two-point correlation function (2PCF) method (e.g., \citealt{Hennawi2021}).
The first approach requires high-quality spectra of individual objects (e.g., \citealt{LiJ2025}) and is therefore not well suited to the current survey design. Tomographic reconstruction (approach 2) demands a very high surface density of background sightlines, while the third class of methods can extract meaningful information from lower-quality spectra, provided the sample size is sufficiently large.

In the following sections, we quantitatively justify the technical requirements of the different observational components adopted in this joint survey. Because the photometric, imaging, and spectroscopic programs probe distinct physical regimes, their sensitivities are naturally expressed in different forms --- e.g., magnitude limits for continuum detection of galaxies and AGN, surface brightness thresholds for diffuse CGM emission, and column density limits for absorption-line measurements of large-scale structures. Rather than directly comparing quoted depth values across facilities, \S~\ref{sec:PhotometricSurvey} and \ref{sec:SpectroscopicSurvey} translate these instrumental limits into physically meaningful detection thresholds relevant to our three science objectives, thereby establishing a coherent sensitivity framework for the combined survey strategy.

\section{Photometric Surveys} \label{sec:PhotometricSurvey}

\subsection{Required source density and detection limit} \label{subsec:SrcDenDetLim}

Mapping the IGM through metal absorption lines requires a dense grid of background sources at redshifts higher than the absorbing structures and brighter than the spectroscopic detection limit. Because the surface density of AGN alone is insufficient to sample the IGM on the relevant spatial scales, we consider both AGN and galaxies as background light sources, following recent justifications of future IGM tomography observations (e.g., \citealt{Japelj2019}).

We estimate the cumulative number of background sources by integrating appropriate luminosity functions (LFs) over comoving volume above a given redshift and down to a specified apparent magnitude limit. For galaxies, we adopt a rest-frame UV ($\sim1500-1700~\text{\AA}$) Schechter LF with redshift-evolving parameters representative of UV-selected star-forming galaxies (e.g., \citealt{Reddy2009, Bouwens2015, Parsa2016}). The LF is parameterized as:
\begin{equation}
\phi(M,z) = 0.4 \ln 10 \, \phi^*(z) \, 10^{0.4 [M^*(z)-M] (\alpha+1)}
    \exp\!\left[-10^{0.4 (M^*(z)-M)}\right],
\end{equation}
where the redshift evolution of the characteristic magnitude and normalization follows the empirical fits of \citet{Reddy2009}:
\begin{equation}
\begin{aligned}
\phi^*(z) &= 1.3\times10^{-3}\ \mathrm{Mpc^{-3}}\ \times\ 10^{-0.27(z-3)}, \\
M^*(z) &= -20.97 + 0.14\,(z-3),
\end{aligned}
\end{equation}
with the faint-end slope fixed at $\alpha = -1.73$. Although the \citet{Reddy2009} parameterization is calibrated primarily at $1.9 < z < 3.4$, its mild redshift evolution is broadly consistent with other UV LFs at $z\sim1-2$ (e.g., \citealt{Bouwens2015, Parsa2016}). This makes it suitable for estimating cumulative source counts in our forward-modeling analysis.

\begin{figure}[hbt]
\centering
\includegraphics[width=0.48\textwidth,trim={0in 0in 0in 0in},clip]{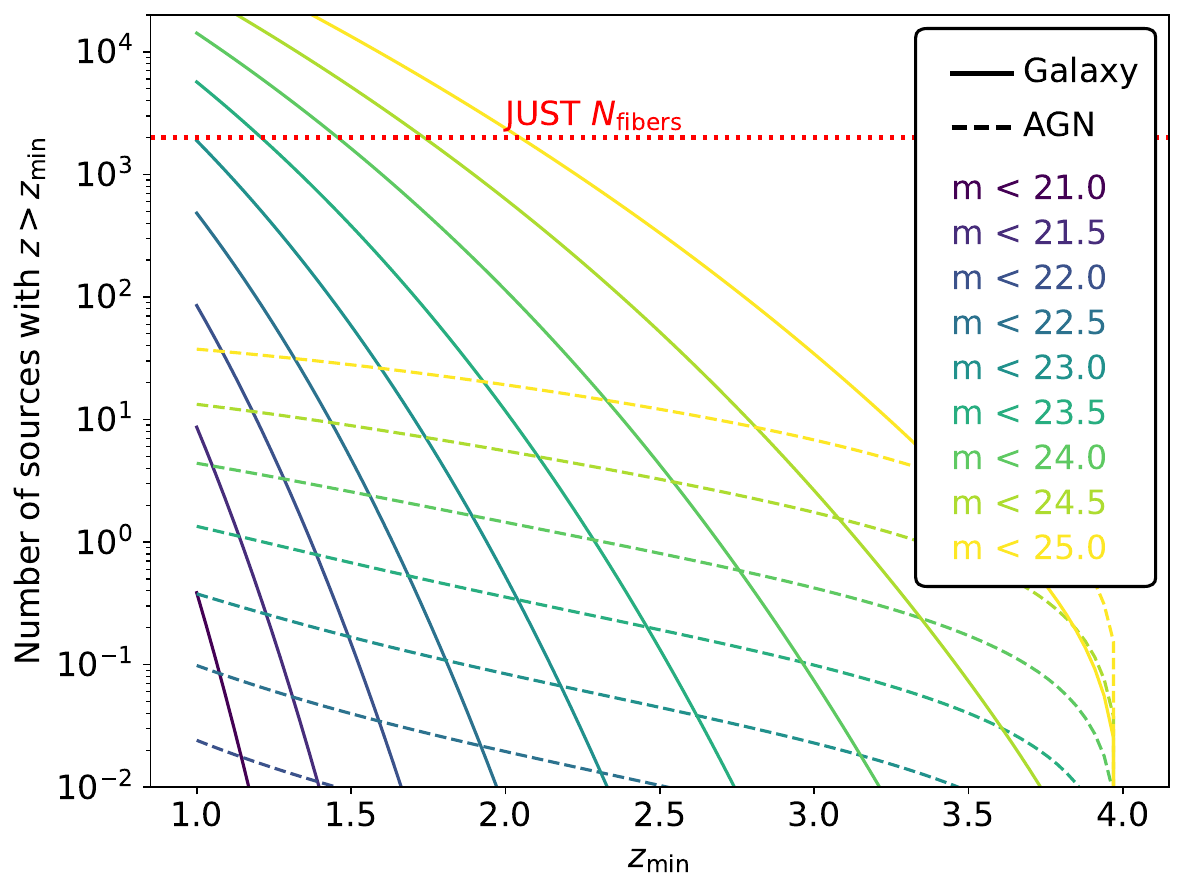}
\caption{Cumulative number of galaxies (solid lines) and AGNs (dashed lines) above a given redshift $z_{\rm min}$ for several magnitude limits ($m < 22.0-25.0$), computed over a $1.4\rm~deg^2$ field corresponding to the FOV of a single JUST pointing. The horizontal dashed line marks the total number of spectroscopic fibers available per pointing ($N_{\rm fibers} = 2000$), indicating the effective upper limit on the number of background sources that can be targeted simultaneously.}\label{fig:NgalaxyAGN}
\end{figure}

\begingroup
\setlength{\tabcolsep}{3pt}
\begin{table*}[htbp]
\centering
\caption{Cumulative number of galaxies and AGNs per JUST field (1.4 deg$^2$) above various redshift thresholds and down to different magnitude limits. The magnitude is generally calculated in rest-frame UV, or falls in the optical band in our redshift of interest.}
\label{tab:source_counts}
\begin{tabular}{c|cc|cc|cc|cc|cc|cc|cc}
\toprule
{$z_\mathrm{min}$} & \multicolumn{2}{c|}{$m < 22.0$} & \multicolumn{2}{c|}{$m < 22.5$} & \multicolumn{2}{c|}{$m < 23.0$} & \multicolumn{2}{c|}{$m < 23.5$} & \multicolumn{2}{c|}{$m < 24.0$} & \multicolumn{2}{c|}{$m < 24.5$} & \multicolumn{2}{c}{$m < 25.0$} \\
 & $N_\mathrm{gal}$ & $N_\mathrm{AGN}$ & $N_\mathrm{gal}$ & $N_\mathrm{AGN}$ & $N_\mathrm{gal}$ & $N_\mathrm{AGN}$ & $N_\mathrm{gal}$ & $N_\mathrm{AGN}$ & $N_\mathrm{gal}$ & $N_\mathrm{AGN}$ & $N_\mathrm{gal}$ & $N_\mathrm{AGN}$ & $N_\mathrm{gal}$ & $N_\mathrm{AGN}$ \\
\midrule
1.0 & 85 & 0 & 483 & 0 & 1887 & 0 & 5685 & 1 & 14225 & 4 & 31088 & 13 & 61413 & 37 \\
1.1 & 23 & 0 & 184 & 0 & 895 & 0 & 3144 & 1 & 8760 & 4 & 20675 & 12 & 43189 & 35 \\
1.2 & 8 & 0 & 84 & 0 & 492 & 0 & 1966 & 1 & 5990 & 4 & 15060 & 11 & 32927 & 33 \\
1.3 & 3 & 0 & 36 & 0 & 261 & 0 & 1202 & 1 & 4036 & 3 & 10866 & 11 & 24948 & 32 \\
1.4 & 0 & 0 & 11 & 0 & 106 & 0 & 601 & 1 & 2329 & 3 & 6926 & 10 & 17064 & 29 \\
1.5 & 0 & 0 & 4 & 0 & 52 & 0 & 348 & 1 & 1514 & 3 & 4883 & 9 & 12738 & 28 \\
1.6 & 0 & 0 & 1 & 0 & 24 & 0 & 196 & 1 & 968 & 2 & 3407 & 8 & 9446 & 26 \\
\bottomrule
\end{tabular}
\end{table*}
\endgroup

For AGN, we adopt the $g$-band LF from \citet{PalanqueDelabrouille2016}, derived from a variability-selected quasar sample with well characterized redshift evolution and a robust faint-end slope. The LF is calibrated over the redshift range $0.68 \lesssim z \lesssim 4.0$, with explicitly modeled evolution in both the normalization and the break magnitude; in our calculations, we restrict the integration to this regime. The LF is described by a double power-law:
\begin{equation}
\phi(M, z) = \frac{\phi^*(z)}
{10^{0.4 \alpha [M - M^*(z)]} + 10^{0.4 \beta [M - M^*(z)]}},
\end{equation}
where the redshift evolution of the normalization and break magnitude follows their best-fit model:
\begin{equation}
\begin{aligned}
\phi^*(z) &= 1.14 \times 10^{-6} 
            \left( \frac{1+z}{3} \right)^{5.0},\\[3pt]
M^*(z)   &= -21.61 
            - 1.37 \log_{10}\!\left( \frac{1+z}{3} \right).
\end{aligned}
\end{equation}
The faint- and bright-end slopes are fixed at $\alpha = -1.3$ and $\beta = -3.2$, respectively.

For each apparent magnitude threshold $m$, we compute the corresponding limiting absolute magnitude at redshift $z$, integrate the LF to obtain the comoving number density of sources brighter than this limit, and then multiply by the comoving volume element. Integrating this quantity from $z_{\rm min}$ to $z_{\rm max}=4$ (the precise choice of $z_{\rm max}$ has negligible impact on the results) over the JUST FOV of $1.4\rm~deg^2$ yields the total number of detectable background galaxies and AGNs. We emphasize that the magnitude thresholds shown in Fig.~\ref{fig:NgalaxyAGN} and Table~\ref{tab:source_counts} represent cumulative source counts within the JUST FOV for a range of illustrative apparent magnitude limits. They do not correspond to the adopted spectroscopic detection limit of a single JUST pointing. The achievable detection limits for the spectroscopic program are quantified separately in \S~\ref{sec:SpectroscopicSurvey}.

To evaluate the feasibility of IGM tomography at $z\sim1.0-1.5$, we compute the number of galaxies and AGNs above several redshift thresholds; the results are summarized in Table~\ref{tab:source_counts}. At any reasonable detection limit, the number of galaxies exceeds that of AGNs by more than at least two orders of magnitude, ensuring that a galaxy-targeted strategy can achieve the high sightline densities required for IGM tomography. The $\sim 2,000$ fibers available within the $\sim 1.4\rm~deg^2$ JUST field are sufficient to observe all available background galaxies in a single pointing for plausible magnitude and redshift cuts (e.g., $m < 23.5\rm~mag$ and $z > 1.3$; see detailed justifications below).

Efficient photometric preselection --- through photo-$z$ or dropout techniques --- will therefore be essential for identifying background galaxies above the desired redshift threshold. Although AGNs are significantly rarer, they can be prioritized when present because their smoother continuum yield higher signal-to-noise ratios (S/N) for detecting IGM absorption features.

\subsection{Photometric surveys used in the analysis}
\label{subsec:ExpPhotometry}

Our photometric data primarily rely on archival Euclid observations of the EDF-N, complemented by CSST and JPCam imaging covering the same sky area. Although future near-IR imaging from facilities such as the Nancy Grace Roman Space Telescope could further enhance photometric depth and redshift accuracy, the survey configuration adopted in this work focuses on Euclid, CSST, and JPCam, which together already provide the wavelength coverage and sensitivity required for our scientific goals. In this section, we justify the adopted detection limits and/or exposure time requirements for these photometric surveys in the context of our scientific goals.


EDF-N is one of Euclid’s three ultra-deep survey fields (the others being the Euclid Deep Field South, EDF-S, and the Euclid Deep Field Fornax, EDF-F). These fields are designed to reach depths of $\gtrsim2\rm~mag$ fainter than the Euclid Wide Survey through repeated observations over relatively small areas \citep{Euclid2022}. EDF-N is centered at $\rm{RA}=17{:}58{:}55.9$, $\rm{Dec}=+66{:}01{:}03.7$, close to the North Ecliptic Pole. This location provides excellent year-round visibility from Euclid’s L2 orbit, as well as low Galactic extinction and minimal zodiacal background. The total area of EDF-N is $\gtrsim20\rm~deg^2$.

Over the nominal $\sim6$-year mission lifetime, the Euclid Deep Fields will accumulate $\sim40$ visits, reaching 5$\sigma$ point-source depths of $I\simeq28.2$ and $Y,J,H\simeq26.4-26.5$ (AB) in the co-added imaging \citep{vanMierlo2022}. These depths are substantially deeper than required for our science goals, which primarily involve selecting background galaxies bright enough for follow-up spectroscopic observations (typically $m \lesssim 24~\mathrm{mag}$; see Fig.~\ref{fig:NgalaxyAGN} and \textsection\,\ref{subsec:SciObj}). At $z\sim1.5$, such deep optical/near-IR imaging reaches well below $L^{\star}$; however, translating a single-band detection limit into a stellar-mass limit is model dependent. For the purposes of this work, it is sufficient to note that the Euclid Deep imaging should probe down to $\sim10^{8.5-9}\rm~M_{\odot}$ for typical blue, star-forming galaxies and to $\sim10^{9.5-10}~M_{\odot}$ for red, quiescent systems at $z\sim1.5$ (order-of-magnitude estimates; the exact completeness limits depend on the adopted selection and SED-fitting assumptions). Reaching well below $L^\star$ ensures that the galaxy population traced by Euclid photometry is representative enough to map large-scale structures such as galaxy groups, (proto-)clusters, and the cosmic web.

In contrast, slitless spectroscopy obtained with the Near-Infrared Spectrometer and Photometer (NISP) is significantly shallower than the broad-band imaging and is usually characterized by a limiting emission line flux \citep{Euclid2023,Euclid2025}. For the Wide survey, Euclid forecasts an unresolved emission line sensitivity of $\sim3\times10^{-16}\rm~erg~s^{-1}~cm^{-2}$ at S/N$=5$. For the deep survey, a significantly better sensitivity of $\sim(0.5-0.8)\times10^{-16}~\mathrm{erg~s^{-1}~cm^{-2}}$ could be reached. At $z\sim1.5$, these limits correspond to characteristic H$\alpha$ SFR thresholds of $\sim$few~$\rm M_{\odot}~yr^{-1}$ for the Deep Fields. While this limits the completeness of emission-line selected samples (especially once dust attenuation is included), it remains adequate for identifying actively star-forming galaxies that preferentially trace overdense environments.


The CSST, also known as Xuntian, is scheduled for launch around 2027 and will operate in conjunction with the Chinese space station, Tiangong. CSST is a 2m aperture near-UV to optical telescope, covering a wavelength range of $\sim2550-10,000~\text{\AA}$, with a wide FOV of $\approx1.1\rm~deg^2$. It is equipped with multiple scientific instruments, including the survey camera for multi-band imaging and slitless spectroscopy, the THz Spectrometer (TS), the Multichannel Imager (MCI), the Integral Field Spectrograph (IFS), and the Cool Planet Imaging Coronagraph (CPI-C) \citep{Zhan2021,CSST2025}.

CSST will carry out several imaging and slitless spectroscopic surveys with different combinations of sky coverage and depth, including seven-band imaging and three-band slitless spectroscopy (e.g., \citealt{Gong2019,Zhan2021,CSST2025}). The planned wide survey will cover approximately $17,500\rm~deg^2$ (about $40\%$ of the sky), reaching a 5$\sigma$ photometric depth of $\sim25.9\rm~mag$ in the $i$ band with $150\rm~s\times2$ exposures, and a slitless spectroscopic depth of $\sim21\rm~mag$ in the GV band ($\sim4000-6500~\text{\AA}$) with $150\rm~s\times 4$ exposures. In addition, CSST is expected to conduct one or more deeper surveys over smaller areas, although their final designs have not yet been fixed. For example, a representative deep-survey configuration with $250\rm~s\times 8$ for imaging and $250\rm~s\times 16$ for spectroscopy would reach detection limits more than one magnitude deeper than those of the wide survey, making them comparable to, or slightly deeper than, the EDF-N in optical.

Compared to Euclid, CSST provides improved coverage at bluer wavelengths, which is particularly important for studies at the cosmic noon. For instance, the Ly$\alpha$ line enters the bluest CSST band at $z\gtrsim1.1$. When combined with archival Euclid data, the CSST imaging and slitless spectroscopy will therefore offer an ideal dataset for selecting galaxies and AGNs with accurate photometric and spectroscopic redshifts at $z\sim1-2$, enabling efficient target selection for follow-up observations.


Broad-band photometry from Euclid and CSST will provide reliable photo-$z$ for identifying background galaxies and AGNs and for tracing large-scale structures. However, broad-band data alone are typically insufficient to robustly identify gravitationally bound systems, such as galaxy groups and (proto-)clusters. This limitation arises because key spectral features, including the $\sim4000~\text{\AA}$ Balmer break and strong nebular emission lines, are only coarsely sampled by broad filters, leading to redshift-color degeneracies and systematic photo-$z$ biases. Additional narrow- or medium-band photometry is therefore essential to achieve the redshift precision required for reliably identifying bound systems in this redshift range.

J-PAS is a wide-field optical survey conducted at the Observatorio Astrof\'{i}sico de Javalambre (OAJ) in Spain, using the 2.5m JST250 and the 1.2~Gpixel JPCam imager \citep{Benitez2014}. The survey is designed to map $\gtrsim8,000\rm~deg^2$ of the northern sky using a set of 56 filters, including 54 contiguous narrow bands with FWHM $\approx145~\text{\AA}$ spanning $\sim3,500-10,000~\text{\AA}$ and spaced by $\approx100~\text{\AA}$, complemented by a small number of broad bands. This configuration effectively provides low-resolution spectrophotometry and enables a typical photo-$z$ accuracy of $\sigma_z \approx 0.003(1+z)$ for bright galaxies (e.g., \citealt{Bonoli2021, HernanCaballero2021}).

Although the EDF-N is not currently included in the nominal J-PAS survey footprint \citep{Benitez2014}, it is observable from the OAJ and could be pursued through the JST250 Open Time program, subject to standard competitive time-allocation procedures. Moreover, given our focus on a relatively narrow redshift interval (e.g., $z\sim1.0-1.4$), the full OAJ/JPCam narrow-band filter set is not required. Instead, a tailored subset of contiguous narrow-band filters that captures a small number of strong spectral features can deliver most of the photo-$z$ improvement. For example, the primary tracer of star-forming galaxies in this redshift range, the [\ion{O}{II}] $\lambda\lambda3727,3729~\text{\AA}$ doublet, would be sampled by $\sim12-13$ narrow-band filters. This typically requires at most two 14-position filter trays (e.g., the current trays 3 \& 4 in the J-PAS observation planner\footnote{\url{https://www.cefca.es/jop/plan/field.html}\label{fn:JPASObsPlanner}}) of JPCam, maximizing observational efficiency.

\subsection{Photometric redshift accuracy}
\label{subsec:Photozaccuracy}


\begin{figure*}
\centering
\includegraphics[width=\linewidth]{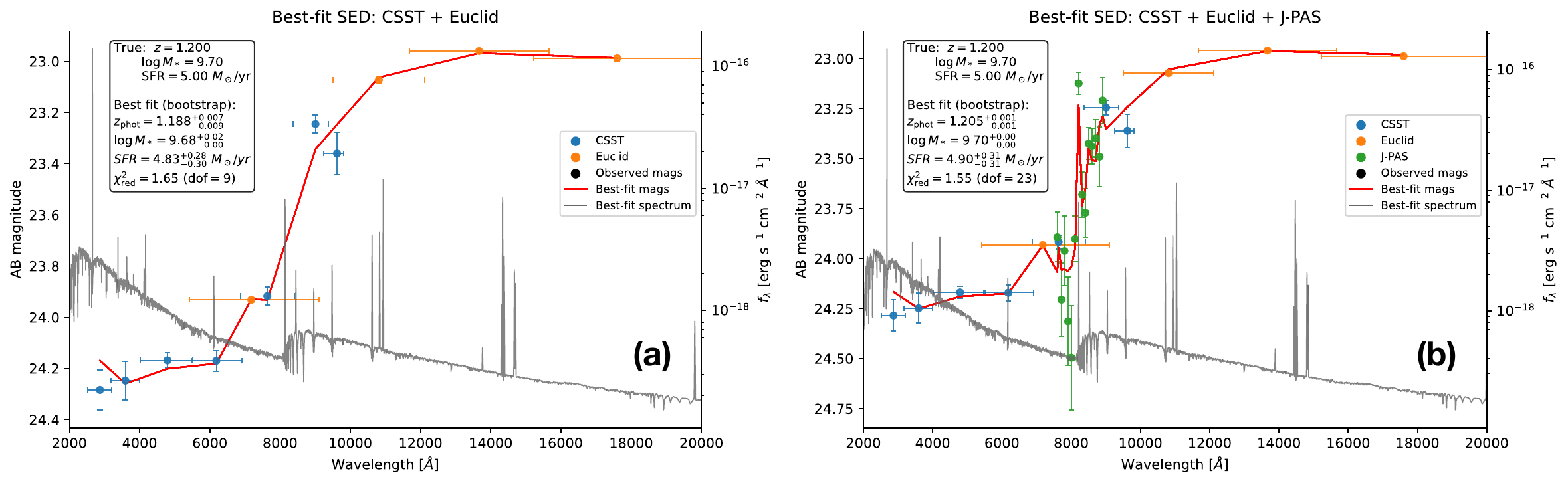}
\caption{Example of the forward modeling SED fitting procedure. The grey curve shows the intrinsic model spectrum generated from our stellar continuum + empirical emission-line prescription (see text in \textsection\,\ref{subsec:ExpPhotometry} for details). Colored points indicate the synthetic photometry with noise added according to the survey detection limits in each filter (CSST, Euclid, and a selected subset of 14 OAJ/JPCam narrow-bands). Horizontal bars show the effective wavelength range contributing to each bandpass. The best-fit photometric SED model, obtained by minimizing $\chi^2$ over redshift, stellar mass, SFR, and dust attenuation, is shown as a solid red curve. The inset text lists the input (``True'') and recovered (``Best fit'') physical parameters, including their $1\sigma$ uncertainties derived from bootstrap resampling of the observed photometry.}\label{fig:sed_fit}
\end{figure*}

In the following, we quantify the improvement in photo-$z$ performance obtained by combining Euclid and CSST data with a realistic subset of JPCam narrow-band filters. Improved photo-$z$ accuracy will be critical not only for the study of large-scale galaxy structures, but also important for large-scale gas structures traced by the Ly$\alpha$ forest or other metal lines (e.g., \citealt{Tan2025}). In our following analysis, the stellar continuum is modeled using the \textsc{Bagpipes} spectral energy distribution (SED) synthesis framework \citep{Carnall2018}. We adopt a single-component exponentially declining star-formation history, parameterized by $(M_\star, Z_\star, t_{\rm age}, \tau)$, and apply dust attenuation following the Calzetti law \citep{Calzetti2000}. To maintain full control over the treatment of emission-line physics, we deliberately disable the internal nebular emission module in \textsc{Bagpipes} and model nebular lines separately.

Nebular emission is added analytically by scaling emission line luminosities to the instantaneous SFR using empirical calibrations. Hydrogen recombination lines are normalized according to the H$\alpha$-SFR relation of \citet{Kennicutt1998,Kennicutt2012},
\begin{equation}
    L_{{\rm H}\alpha} = 1.8\times10^{41}
    \left( \frac{{\rm SFR}}{\rm M_\odot\,yr^{-1}} \right)
    ~{\rm erg~s^{-1}},
\end{equation}
while higher-order Balmer and Paschen lines are derived assuming Case~B recombination ratios \citep{Osterbrock2006}. 

We further include all major rest-frame UV and optical metal lines relevant for $2000-20,000~\text{\AA}$ photometry at $z\sim1-2$. These include [\ion{O}{II}] $\lambda\lambda3727,3729~\text{\AA}$, [\ion{O}{III}] $\lambda\lambda4959,5007~\text{\AA}$, [\ion{Ne}{III}] $\lambda\lambda3869,3968~\text{\AA}$, [\ion{N}{II}] $\lambda\lambda6548,6583~\text{\AA}$, [\ion{S}{II}] $\lambda\lambda6716,6731~\text{\AA}$, [\ion{S}{III}] $\lambda\lambda9069,9531~\text{\AA}$, Ly$\alpha$ $\lambda1216~\text{\AA}$, \ion{C}{III}] $\lambda\lambda1907,1909~\text{\AA}$, \ion{C}{IV} $\lambda\lambda1548,1550~\text{\AA}$, \ion{He}{II} $\lambda1640~\text{\AA}$, and \ion{Mg}{II} $\lambda\lambda2796,2803~\text{\AA}$. Relative metal-line strengths are adopted from empirical calibrations appropriate for typical star-forming galaxies at $z\sim1-2$ \citep{Kewley2002,Moustakas2006,Gutkin2016}. All emission lines are modeled with Gaussian profiles with a full width at half maximum of ${\rm FWHM}=250~{\rm km\,s^{-1}}$, representative of nebular linewidths in main-sequence galaxies at these redshifts.

We obtain the filter transmission curves for CSST, Euclid, and OAJ/JPCam through the SVO Filter Profile Service \citep{Rodrigo2020}. For each model SED, we compute noiseless AB magnitudes through all filters via standard photon-weighted integration.

\begin{figure*}
\centering
\includegraphics[width=\textwidth]{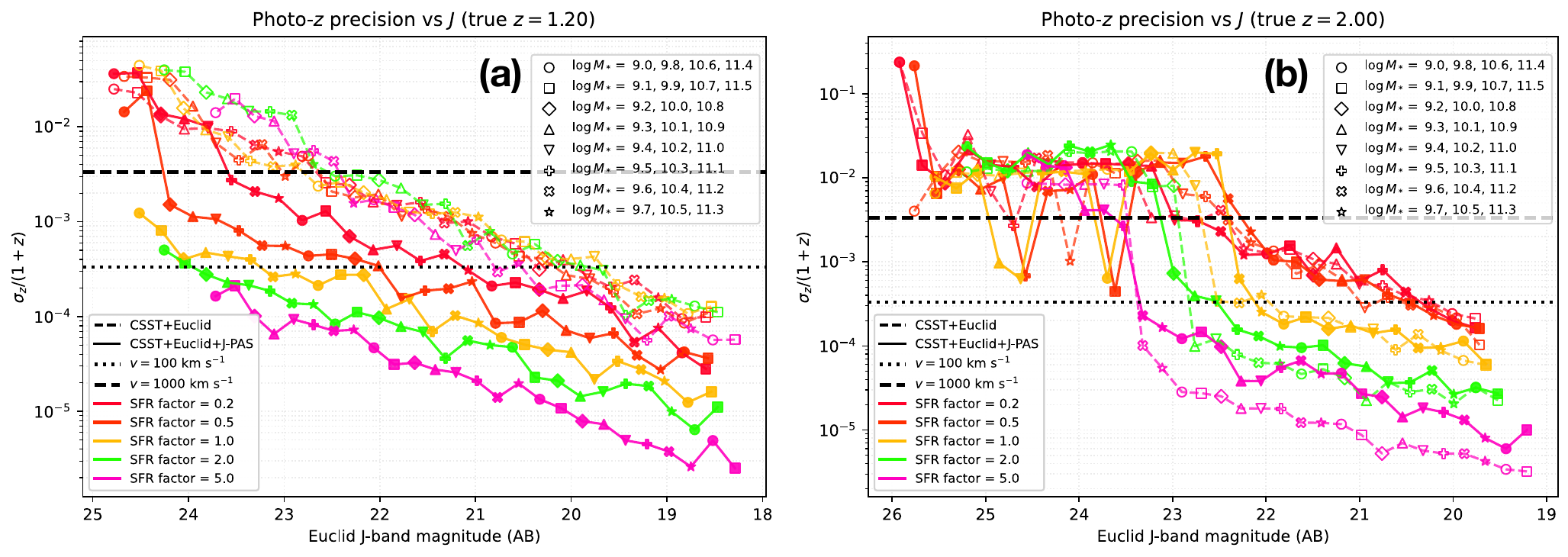}
\caption{Photometric redshift precision as a function of Euclid $J$-band magnitude for galaxies at two fiducial redshifts ($z_{\rm true}=1.2$ and 2.0). In each panel, the vertical axis shows the redshift uncertainty $\sigma_z/(1+z)$ in logarithmic scale, derived from bootstrap resampling of noisy synthetic photometry. Colors represent different SFR scaling factors relative to the star-forming main sequence, while marker shapes encode different stellar masses. Dashed lines correspond to fits using only broad-band data (CSST + Euclid), whereas solid lines include the 14 JPCam narrow-band filters that bracket redshifted [\ion{O}{II}] and the Balmer break at $z\sim1.0-1.4$. The two horizontal lines denote the corresponding velocity measurement accuracy of $v=100\rm~km~s^{-1}$ and $v=1000\rm~km~s^{-1}$.}\label{fig:photoz_accuracy}
\end{figure*}

To evaluate sensitivity to survey depth, we generate observed photometry by adding Gaussian noise corresponding to the $n\sigma$ detection limit $m_{\rm lim}$ for each band. Given a limiting flux density $f_{\rm lim}$ at S/N$=n$, the flux and magnitude uncertainties are ($f$ is the measured flux density):
\begin{equation}
    \sigma_f = \frac{f_{\rm lim}}{n}, \qquad
    \sigma_m = \frac{2.5}{\ln 10}\,\frac{\sigma_f}{f}.
\end{equation}
We explore a range of limiting magnitudes for the JPCam bands, corresponding to realistic depths for narrow-band observations. Assuming an average per band exposure of $10\rm~hours$, using the JPCam exposure time calculator, we obtain the detection limits will be typically 23.9-24.6~mag from J0760 to J0890 (OAJ filters used for JPCam; wavelength range $\sim7510~\text{\AA}-9010~\text{\AA}$, typical bandwidth $\sim145~\text{\AA}$; adopted in Figs.~\ref{fig:sed_fit} and \ref{fig:photoz_accuracy}).  

For a grid of redshift $z$, stellar mass $M_\star$, SFR, and dust attenuation $A_V$, we compute synthetic photometry and evaluate the likelihood via: 
\begin{equation}
    \chi^2 (z, M_\star, \mathrm{SFR}, A_V)
    = \sum_{i}
    \left[
        \frac{
            m_{i,\rm obs} - m_{i,\rm model}(z,M_\star,\mathrm{SFR},A_V)
        }{
            \sigma_{m,i}
        }
    \right]^2,
\end{equation}
where the sum extends over all filters with detections. The best-fit redshift is obtained by minimizing $\chi^2$ across the parameter grid (e.g., Fig.~\ref{fig:sed_fit}). To assess uncertainties, we perform $N_{\rm boot}$ bootstrap realizations in which the observed photometry is regenerated with new noise draws. This gives a robust estimate of the achievable photo-$z$ precision at a given survey depth.

In both our photo-$z$ and later the CGM emission (\textsection\,\ref{subsec:CGMNarrowBandImaging}) justifications we parametrize the SF activity of galaxies in terms of a multiplicative factor relative to the star-forming main sequence, $f_{\rm SFR}\equiv{\rm SFR}/{\rm SFR_{MS}}$. For concreteness, we adopt a simple analytic form for the main sequence that is tuned to reproduce the redshift and mass dependence compiled by \citet{Speagle2014} at $z\sim1-2$:
\begin{equation}\label{eq:ms_param}
  \mathrm{SFR}_{\rm MS}(M_\star,z)
  \;=\;
  10~\mathrm{M_\odot~yr^{-1}}\,
  \left(\frac{M_\star}{10^{10}~\mathrm{M_\odot}}\right)^{0.8}
  \left(\frac{1+z}{2}\right)^{2.5},
\end{equation}
which yields a slope in stellar mass and a redshift evolution broadly consistent with observational determinations of the main sequence at cosmic noon (e.g., \citealp{Speagle2014,Whitaker2014,Schreiber2015}). Eq.~\ref{eq:ms_param} is not intended as a new fit, but as a convenient parametrization within the envelope of these measurements.

We evaluate the photo-$z$ precision across a realistic galaxy parameter space: stellar masses $\log(M_\star/M_\odot)=9-11.5$, SFR factor of $f_{\rm SFR}=0.2-5$. The corresponding resultant Euclid $J$-band magnitude is typically $\sim18-26\rm~mag$ (Fig.~\ref{fig:photoz_accuracy}). Two photometric configurations are compared:
\begin{enumerate}
    \item Broad bands only: CSST + Euclid.
    \item Broad + narrow: CSST + Euclid + 14 JPCam bands, chosen to bracket the redshifted [\ion{O}{II}] and Balmer break at $z\sim1.2$ (narrow-band filters from J0760 to J0890; Fig.~\ref{fig:sed_fit}).
\end{enumerate}

As shown in Fig.~\ref{fig:sed_fit}, in our redshift of interest ($z\sim1-2$), the strongest feature to constrain the redshift, the Balmer break, falls roughly at the reddest bands of CSST and the bluest bands of Euclid. Therefore, combining they two will lead to a good photometric redshift accuracy of $\Delta z\sim0.01$ ($\Delta v\gtrsim10^3\rm~km~s^{-1}$ in our redshift range of interest). This is typically good enough to identify large-scale structures and background galaxies. However, these broad-band photometries are insensitive to strong emission lines from star-forming galaxies, limiting the photo-$z$ accuracy to identify gravitationally bound systems. As further justified in Fig.~\ref{fig:photoz_accuracy}, adding narrow-band photometry from JPCam could significantly improve the photo-$z$ measurement accuracy, although the detection limit is typically significantly higher than space observations of CSST and Euclid. In appropriate redshift range (e.g., Fig.~\ref{fig:photoz_accuracy}a at $z=1.2$), where the strong [\ion{O}{II}] line falls in the adopted JPCam filters, the photo-$z$ accuracy could be typically about one order of magnitude higher than those obtained from pure broad band SED. The improvement is more significant when the SFR is higher, as the relative strength of the emission lines to continuum is higher. The photo-$z$ accuracy also strongly depends on the brightness of the galaxy, which in turn mostly depends on the stellar mass, but the relative improvement by adding the narrow-band data does not strongly depend on the magnitude. Assuming we need a velocity measurement accuracy of $v\sim300\rm~km~s^{-1}$ to identify galaxy groups or (proto-)clusters, under the current adopted survey depth (EDF-N for Euclid, deep survey for CSST, and $\sim10\rm~hours$ exposure for JST250/JPCam), this accuracy can be achieved for galaxies with a $J$-band magnitude of $\sim24\rm~mag$ (for main sequence SFR), close to the requested narrow-band detection limits of JPCam. Without the narrow-band data from JPCam, such an accuracy can only be achieved for galaxies brighter than $\sim21\rm~mag$, which are far less abundant at $z\sim1-2$ (Fig.~\ref{fig:NgalaxyAGN}). The photo-$z$ accuracy improvement will be quite limited at $z\gtrsim1.5$ (e.g., Fig.~\ref{fig:photoz_accuracy}b), where the [\ion{O}{II}] line and sometimes also the Balmer break, move out of the wavelength range covered by the selected narrow-band filters. 

Medium- and narrow-band photometry have long been recognized as powerful tools for improving photo-$z$ accuracy beyond what is achievable with broad bands alone. Our results, based on controlled forward modeling with realistic SEDs, confirm the importance of improving the photo-$z$ accuracy by adding only a few narrow-band filters covering special spectral features (Balmer break and [\ion{O}{II}]) in a certain redshift range ($z\sim1.0-1.4$). The survey, based on deep space-based observations from Euclid and CSST, and dedicated narrow-band imaging from JPCam, is quite competitive compared to similar existing surveys for its depth and/or photo-$z$ accuracy. 

In addition to the full J-PAS survey \citep{Benitez2014}, there are some other photo-$z$ surveys employing medium- or narrow-band filters which often covers significantly smaller sky areas. The COSMOS survey (e.g., \citealt{Ilbert2009,Laigle2016}), which employs a combination of broad- and medium-band filters from a few ground-based telescopes including Subaru, achieves typical photo-$z$ uncertainties of $\sigma_z/(1+z) \sim 0.007-0.03$ for $i<24$ galaxies at $z<1.2$. However, at higher redshift, the COSMOS medium-band coverage becomes sparse, and the precision degrades significantly due to the limited sampling of strong rest-frame optical emission lines and the Balmer break. The ALHAMBRA survey \citep{Moles2008, Molino2014}, conducted with the Calar Alto 3.5m telescope, demonstrated that $\sim20$ optical medium-band filters spanning $3500-9700~\text{\AA}$ plus some broad near-IR filters can achieve $\sigma_z/(1+z)\lesssim0.03$ at a detection limit comparable to or deeper than our survey. The PAU Survey (PAUS) \citep{Eriksen2019} uses the 4.2m William Herschel Telescope and 40 narrow-band filters similar as those used on J-PAS, achieving $\sigma_z/(1+z) \sim 0.003$ for bright galaxies ($i<22.5\rm~mag$) and an even higher accuracy of $\sigma_z/(1+z) \sim 0.001$ for brighter emission line galaxies at $z<1.2$.  

Overall, our analysis demonstrate that medium- or narrow-band photometry need not cover the entire optical range uniformly to deliver substantial photo-$z$ accuracy improvement to identify gravitationally bound systems. Strategically placed bands that sample key emission lines and continuum breaks at the redshift of interest can achieve performance comparable to or better than existing medium- or narrow-band surveys with many more filters. In the meantime, the broad-band space observations guarantee the survey depth to achieve our goal of identifying background galaxies.

\subsection{CGM detection based on JPCam narrow-band imaging} \label{subsec:CGMNarrowBandImaging}

As outlined in \textsection\,\ref{subsec:SciObj}, an important goal of our program is to map the diffuse CGM in emission during the cosmic noon. While most existing narrow-band studies of extended halos have targeted Ly$\alpha$ at higher redshifts (e.g., \citealt{Cantalupo2014,Borisova2016,Cai2017,ArrigoniBattaia2019}), our strategy instead focuses on rest-frame optical/UV metal lines at $z\sim1-2$, in particular the [\ion{O}{II}] $\lambda\lambda3727,3729~\text{\AA}$ doublet, which traces ongoing SF and metal-enriched gas (e.g., \citealt{Stroe2017,Rupke2019,Dutta2023,Dutta2024}). The contiguous 14 narrow-band filters we selected from OAJ/JPCam provide nearly continuous wavelength coverage over $\sim7510-9010~\text{\AA}$ (the actual redshift/wavelength coverages could be broader if we select two filter trays, based on the required configuration of the J-PAS observation planner$^\text{\ref{fn:JPASObsPlanner}}$), enabling efficient searches for extended [\ion{O}{II}] emission over redshift intervals of $\Delta z \sim 0.04$ per filter and a continuous coverage over $z\sim1.0-1.4$. Other emission lines which could be detected from the extended CGM, such as \ion{Mg}{II} (e.g., \citealt{Guo2023}), could be studied in a similar approach at different redshifts with the same dataset.

\begin{figure}
\centering
\includegraphics[width=0.49\textwidth]{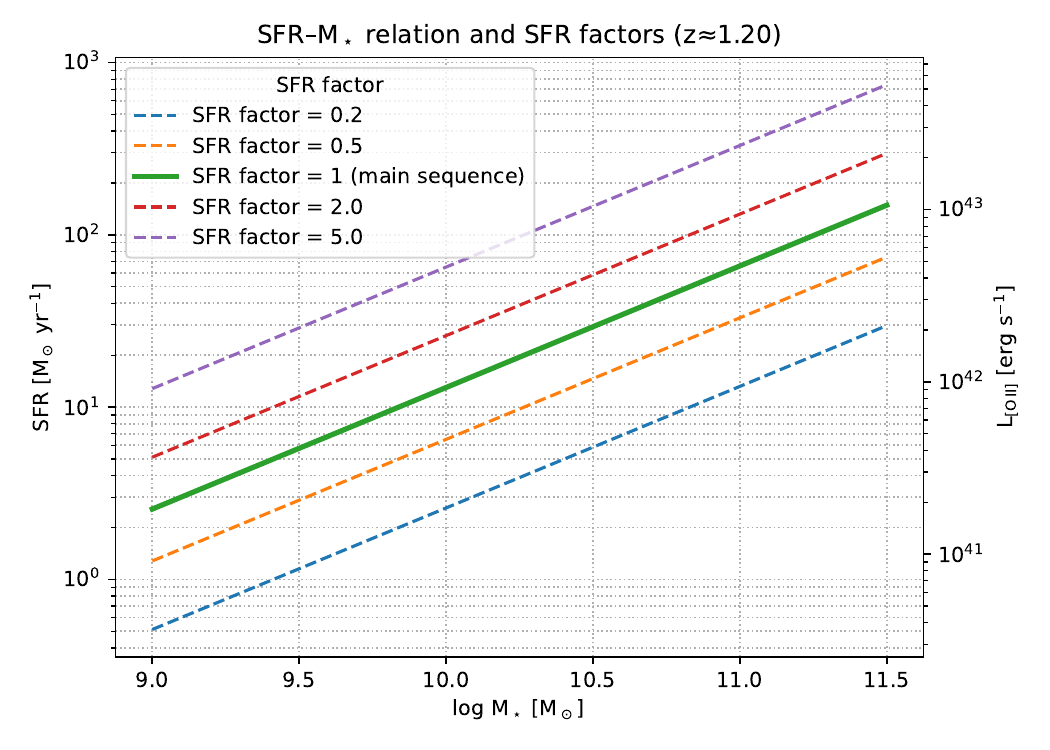}
\caption{Star-forming main sequence and SFR factor tracks at $z\simeq1.2$. The solid line shows the SF main sequence with the SFR factor $f_{\rm SFR}=\mathrm{SFR}/\mathrm{SFR_{MS}}=1$, while dashed curves correspond to SFR factors $f_{\rm SFR}=0.2,\,0.5,\,2,$ and $5$. The left axis gives the SFR, and the right axis shows the corresponding [\ion{O}{II}] luminosity $L_{\rm [O\,II]}$ linearly scaled to the SFR via Eq.~\ref{eq:OIISFR}.}\label{fig:SFRmsOII}
\end{figure}

In \textsection\,\ref{subsec:ExpPhotometry}, we constructed realistic SEDs of galaxies using Bagpipes, sampling stellar masses in the range $\log(M_\star/\mathrm{M}_\odot)\sim9-11.5$ and parameterizing their SFRs in terms of a multiplicative factor $f_{\rm SFR}$ relative to the SF main sequence, with $f_{\rm SFR}=0.2-5$. These models, combined with the galaxy LFs adopted in \textsection\,\ref{subsec:SrcDenDetLim}, define a representative population of star-forming galaxies at the cosmic noon that will be detected in the Euclid+CSST+JPCam photometric data. For the purposes of CGM emission, we convert SFR to [\ion{O}{II}] luminosity via a standard linear calibration,
\begin{equation}\label{eq:OIISFR}
  L_{\rm [O\,II]} \;=\; k_{\rm [O\,II]}\,\mathrm{SFR}\,,
\end{equation}
with $k_{\rm [O\,II]}\sim7\times10^{40}~\mathrm{erg~s^{-1}}/(\mathrm{M_\odot~yr^{-1}})$, consistent with empirical relations for star-forming galaxies at similar redshifts (e.g., \citealt{Kewley2004,Hayashi2013}). This prescription links the SFR and stellar mass distributions constrained by our photo-$z$ experiments directly to a distribution of [\ion{O}{II}] luminosities (Fig.~\ref{fig:SFRmsOII}). Here we do not consider the ionization photons from the UV background, which can make significant contributions to ionize cool gas at $z\sim1$ (e.g., \citealt{Dutta2024}). 

\begin{figure*}
\centering
\includegraphics[width=\textwidth]{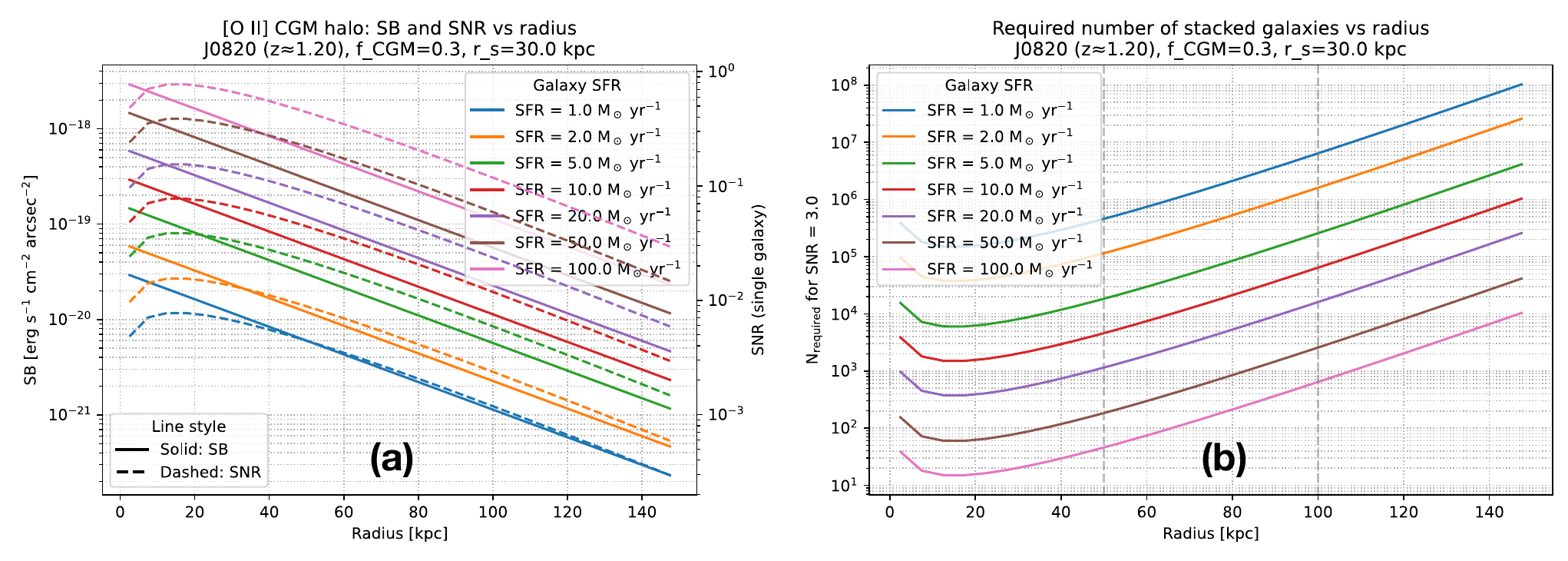}
\caption{Predicted detectability of extended [\ion{O}{II}] CGM emission in JPCam narrow-band imaging at $z\simeq1.2$. (a) Radial [\ion{O}{II}] surface brightness profiles (solid curves; left axis) and the corresponding single galaxy S/N (dashed curves; right axis) for galaxies with $\mathrm{SFR}=1,\,2,\,5,\,10,\,20,\,50,$ and $100~\mathrm{M_\odot~yr^{-1}}$. We adopt the JPCam J0820 filter ($\sim10\rm~h$ exposure, $m_{\rm AB,lim}\simeq24.6$ at $\rm S/N\simeq5$), a [\ion{O}{II}]-SFR conversion using Eq.~\ref{eq:OIISFR}, and simply assume that a fraction $f_{\rm CGM}=0.3$ of the total [\ion{O}{II}] luminosity emerges in an exponential CGM with scale radius $r_{\rm s}=30~\mathrm{kpc}$. A PSF of $\sim1^{\prime\prime}$ has been adopted. Both axes are shown on a logarithmic scale. (b) Number of galaxies required to stack, $N_{\rm req}(r)$, in order to reach a target $\mathrm{S/N}=3$ in the azimuthally averaged [\ion{O}{II}] radial profile, for the same set of SFRs. Vertical dashed lines mark $r=50$ and $100~\mathrm{kpc}$, respectively.}\label{fig:CGMprofile}
\end{figure*}

To model the extended CGM emission, we assume that a fraction $f_{\rm CGM}$ of the total [\ion{O}{II}] luminosity emerges in a diffuse halo, while the remainder comes from the central compact star-forming regions. Motivated by observations of extended halos around star-forming galaxies and quasars (e.g., \citealt{Stroe2017,Rupke2019,Dutta2023,Dutta2024}), we simply adopt a fiducial $f_{\rm CGM}=0.3$ and an exponential surface-brightness profile:
\begin{equation}
  \Sigma_{L}(r) \;=\;
  \frac{L_{\rm CGM}}{2\pi r_{\rm s}^2}
  \exp\!\left(-\frac{r}{r_{\rm s}}\right)\!,
\end{equation}
where $L_{\rm CGM}=f_{\rm CGM}\,L_{\rm [O\,II]}$ and $r_{\rm s}=30~\mathrm{kpc}$ is the halo scale length. This simple parameterization captures the generic decline of surface brightness with galactocentric radius and allows us to predict the observable [\ion{O}{II}] profile for any given combination of ($M_\star$, SFR).

We emphasize that this parameterization is not intended to reproduce the detailed [\ion{O}{II}] surface-brightness distribution in the central star-forming disk, as resolved by integral-field spectroscopy of individual bright systems (e.g., \citealt{Rupke2019,Nielsen2024,Zhang2024}). Instead, our goal is to estimate the detectability of the extended, diffuse CGM component in a statistical narrow-band imaging survey. The adopted model therefore provides a simplified, physically motivated description of the halo component, sufficient for evaluating stacking requirements, without attempting to capture the complex central emission structures that are typically observed in IFS-selected, individually bright galaxies.

We then propagate these model halos through the JPCam filter response. For the subset of 14 filters from J0760 to J0890 that bracket the redshifted [\ion{O}{II}] line and Balmer break at $z\sim1.0-1.4$ (\textsection\,\ref{subsec:ExpPhotometry}), we adopt a representative $\sim10\rm~h$ exposure with 5$\sigma$ point source detection limits of $m_{\rm AB}\approx23.9-24.6$. These depths are converted into a 1$\sigma$ flux uncertainty in a $\sim1^{\prime\prime}$ PSF aperture and, by scaling with the annular area, into a surface brightness noise level in units of $\rm{erg~s^{-1}~cm^{-2}~arcsec^{-2}}$ as a function of radius. For a fiducial redshift $z\simeq1.2$ (corresponding to [\ion{O}{II}] being covered by the J0820 filter), we compute the radial [\ion{O}{II}] surface brightness profiles and the S/N per galaxy in concentric annuli. As shown in Fig.~\ref{fig:CGMprofile}a, for a single galaxy, even when the SFR is as high as $\gtrsim100~\mathrm{M_\odot~yr^{-1}}$, corresponding to a main-sequence galaxy with a stellar mass $M_\star\sim10^{11.1}\rm~M_\odot$ (Fig.~\ref{fig:SFRmsOII}; or $J$-band magnitude of $\sim19.5\rm~mag$, Fig.~\ref{fig:photoz_accuracy}a), we still cannot detect the extended [\ion{O}{II}] emission at a reasonably high S/N. This estimate indicates that only in some extreme cases (when a galaxy has extreme starbursts or hosts an UV bright AGN) could we detect the extended [\ion{O}{II}] emitting CGM around individual galaxies via the proposed JPCam narrow-band imaging. We typically need stacking to study the CGM distribution.

We next justify the feasibility of such stacking analysis based on the galaxy number counts derived in \textsection\,\ref{subsec:SrcDenDetLim} (Fig.~\ref{fig:NgalaxyAGN} and Table~\ref{tab:source_counts}; based on the UV-selected galaxy LF from \citealt{Reddy2009,Bouwens2015}). Assuming that noise is dominated by the sky background and is therefore uncorrelated between galaxies, the S/N in a stacked narrow-band image of $N$ similar systems scales as $\rm{S/N}_{\rm stack}\propto\sqrt{N}$. For each choice of SFR, we invert this relation to estimate the number of galaxies required to reach a target $\rm S/N\approx3$ as a function of radius (Fig.~\ref{fig:CGMprofile}b). For example, here we consider the case of moderate main sequence systems with $\rm{SFR}\sim10~\mathrm{M_\odot~yr^{-1}}$ and $M_\star\sim10^{9.8}\rm~M_\odot$ (Fig.~\ref{fig:SFRmsOII}). Considering the dust attenuation, we simply assume the rest-frame UV magnitude adopted in Fig.~\ref{fig:NgalaxyAGN} is $\sim 1\rm~mag$ fainter than the $J$-band magnitude adopted in Fig.~\ref{fig:photoz_accuracy}a. Therefore, the galaxy will have a UV magnitude of $\approx23.7\rm~mag$, so there will be a few thousand such galaxies at $z\sim1.2$ in one JUST FOV ($\sim1.4\rm~deg^2$, $\sim1/15$ of the survey area). On the other hand, our fiducial halo model implies that detecting the average [\ion{O}{II}] halo at $r\sim50~\mathrm{kpc}$ around such galaxies requires stacking of order $N\sim4\times10^3$ galaxies, while extending the detection to $r\sim100~\mathrm{kpc}$ demands of order $\sim6\times10^4$ galaxies. Considering the uncertainties, we therefore confirm that for the whole survey, we can easily explore the [\ion{O}{II}] emission distribution in the CGM out to $r\gtrsim50~\mathrm{kpc}$ over multiple bins in stellar mass, SFR factor, and other parameters such as the environment (e.g., group/cluster membership). We could also detect the CGM out to $r\gtrsim100~\mathrm{kpc}$ for the entire sample and maybe for a few bins in galaxy parameters. More extreme starbursts with $\rm{SFR}\sim(50-100)\rm~M_\odot~yr^{-1}$ require only $\sim10^2$ galaxies to reach $\rm S/N\approx3$ at $r\sim50\rm~kpc$ and of order $\sim10^3$ at $r\sim100\rm~kpc$, whereas quiescent galaxies with $\rm{SFR}\lesssim1\rm~M_\odot~yr^{-1}$ would require much larger samples and are primarily accessible via very large stacked ensembles in bigger surveys.

We note that the stacking analysis discussed here refers to azimuthally averaged narrow-band imaging. The dominant noise source in this case is expected to be sky background fluctuations, which are largely uncorrelated between galaxies and therefore scale approximately as $\sqrt{N}$. Potential systematics such as residual background subtraction errors or contamination from outliers can be mitigated through careful masking and robust stacking procedures (e.g., median stacking or sigma-clipping). Given the small angular size of the CGM relative to the FOV and the narrow-band nature of the imaging, we expect these effects to be sub-dominant for the order-of-magnitude detectability estimates presented here.

In summary, our modeling demonstrates that JPCam narrow-band imaging, when combined with Euclid and CSST photometry, will enable a statistically powerful census of [\ion{O}{II}] CGM emission at the cosmic noon. By stacking galaxies in different subsamples, we will be able to measure the average radial [\ion{O}{II}] surface brightness profiles out to tens of kpc and, in favorable cases, $\sim100~\mathrm{kpc}$. These measurements will complement the absorption line studies of the CGM/IGM from CSST and JUST which are typically better for larger scale structures (\textsection\,\ref{sec:SpectroscopicSurvey}), providing a self-contained, multi-phase view of how star-forming galaxies exchange mass, metals, and energy with their gaseous environments during the epoch of peak cosmic SF.

\section{Spectroscopy Surveys} \label{sec:SpectroscopicSurvey}

In this section, we justify the IGM tomography science based on absorption lines in the background source (mostly galaxies) spectra. We will mainly use two sets of data: the Ly$\alpha$ absorption trough from the CSST GU band slitless spectra and the metal absorption lines from the JUST multi-object spectra. 

\subsection{Mock three-dimensional gas distribution}\label{subsec:MockGasCube}

In order to test how well background source absorption spectroscopy can recover the cosmic gas distribution, we construct a mock three-dimensional (3D) model of the IGM in our primary redshift range of interest $z=1.0-1.4$ (could be different for different absorption lines, here just take this range as an example). Instead of extracting the 3D gas distribution from any certain numerical simulations, we represent the large-scale gas distribution by a simple smoothed Gaussian random field with a controlled comoving correlation length $L_{\rm corr}$ (we adopt $L_{\rm corr} \simeq 2.5\rm~Mpc$). The procedure is conceptually similar to the Ly$\alpha$ forest tomography simulations of \citet{Japelj2019}, but adapted to our redshift range and to the geometry and depth of the EDF-N. 

The procedure is described in detail in the appendix (\textsection\,\ref{Appsec:3DGasCube}). When generating the mock gas distribution, we adopt a $\Lambda$CDM cosmology with parameters listed in \textsection\,\ref{sec:Introduction}. The FOV in the transverse directions is chosen to match the area of Euclid’s EDF-N, which is planned to cover $\simeq 20~{\rm deg}^2$ \citep{Euclid2022}. For simplicity we approximate EDF-N as a square with an angular size of $4.47^\circ\times4.47^\circ$. At $z_{\rm mid}\simeq1.2$ this corresponds to a transverse comoving size of $\simeq 135\times135\rm~cMpc$. We adopt a resolution in the transverse direction of $512\times512$, so the comoving voxel size in the $(x,y)$ plane is $\Delta x=\Delta y\simeq0.26\rm~cMpc\simeq 120\rm~pkpc$ at $z_{\rm mid}\simeq1.2$. This is typically sufficient to resolve large-scale structures (e.g., cosmic webs and voids), and also more than sufficient for the sampling density by our background galaxies (see justifications below). Gas from smaller scale structures, such as the CGM of individual galaxies, could be studied via the emission line mapping method discussed in \textsection\,\ref{subsec:CGMNarrowBandImaging}.

Along the line of sight (LOS), we choose a resolution of 2048 slices over a redshift range of $z=1.0-1.4$. This corresponds to a resolution of $\Delta z\simeq2\times10^{-4}$, or $\Delta v\simeq 27\rm~km~s^{-1}$ at $z=1.2$. This resolution is high enough compared to our highest resolution spectroscopy observations from JUST ($R\sim4,000-5,000$; \citealt{JUST2024}; $\Delta v\gtrsim 60\rm~km~s^{-1}$), allowing an over sampling of the simulated JUST spectra. The corresponding LOS spacing resolution is $\Delta \chi\simeq0.3\rm~cMpc\simeq137\rm~pkpc$ at $z=1.2$, comparable to the voxel size in the transverse direction. 

\begin{figure*}
  \centering
  \includegraphics[width=1.0\textwidth]{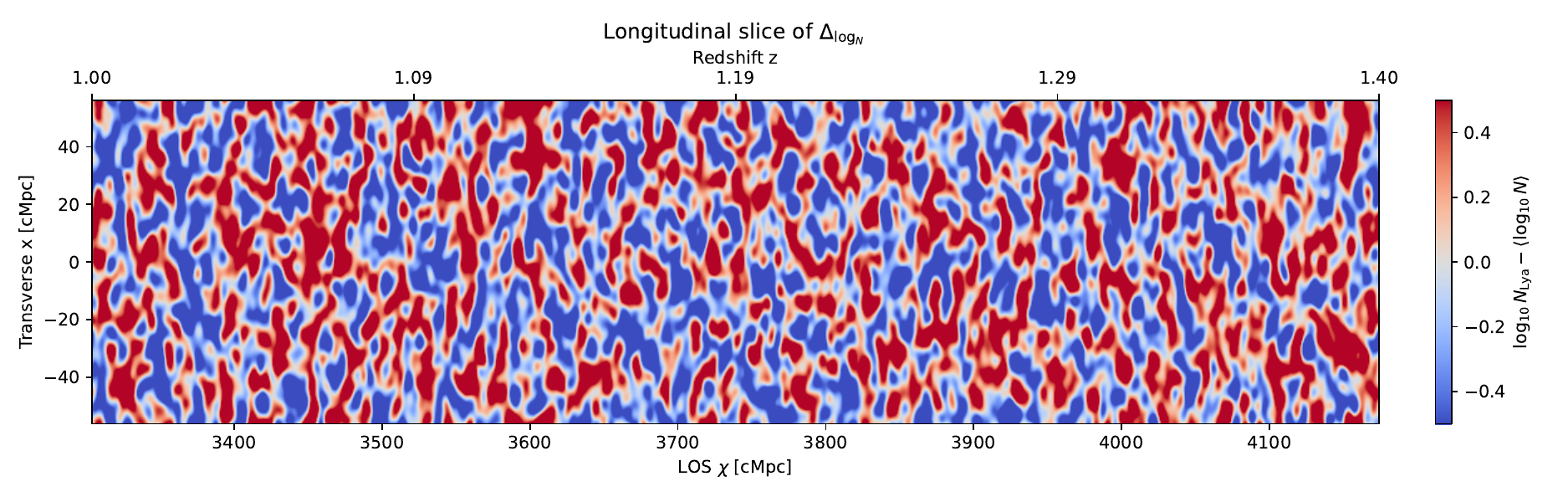}
  \caption{Diagnostic longitudinal slice in the $\chi$ (or $z$)--$x$ plane through the simulated Ly$\alpha$ column density contrast $\Delta_{\log N}$ (Eq.~\ref{eq:Lyaconstrast}) cube. This slice is extracted from the 3D $\Delta_{\log N}$ data cube by averaging over a finite band (10 cMpc) in the transverse $y$ direction. It is plotted as a function of redshift (horizontal axis) and transverse coordinate $x$ (vertical axis).}\label{fig:NHIsliceszx}
\end{figure*}

\begin{figure}
  \centering
  \includegraphics[width=0.49\textwidth]{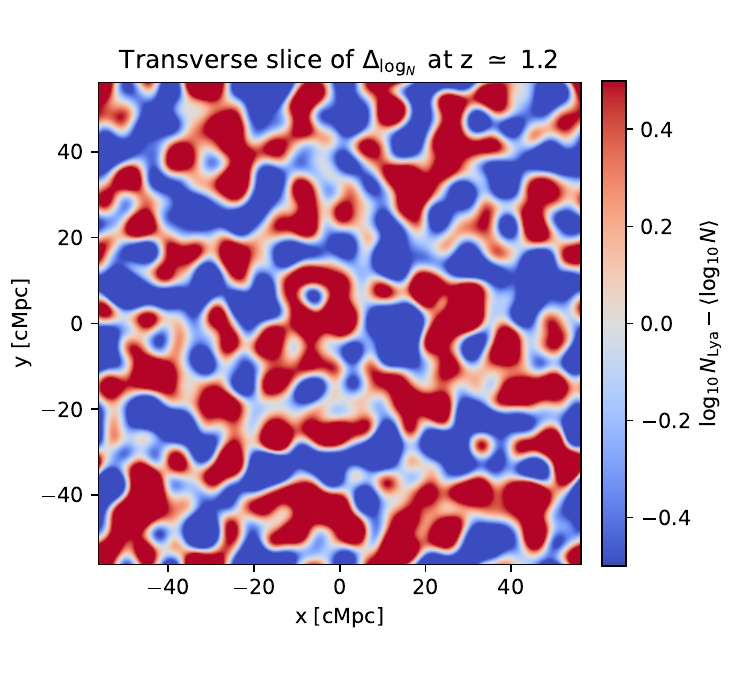}
  \caption{Similar as Fig.~\ref{fig:NHIsliceszx}, but showing a transverse slice of $\Delta_{\log N}$ at $z\simeq 1.2$ in the $x$--$y$ plane.}\label{fig:NHIslicesxy}
\end{figure}

The resulting 3D data cube provides both Ly$\alpha$ optical depth $\tau$ and neutral hydrogen column density $N_{\rm HI}$ as functions of position [$\chi$ in the LOS and $(x, y)$ in the transverse directions, respectively]. To visually inspect the simulated large-scale structures, in Figs.~\ref{fig:NHIsliceszx} and \ref{fig:NHIslicesxy}, we present examples of 2D gas distribution slices along the longitudinal and transverse directions, in unit of $N_{\rm HI}$, extracted from the 3D data cube. The $\tau$ maps are presented in the appendix (\textsection\,\ref{Appsec:3DGasCube}; Figs.~\ref{fig:tausliceszx}, \ref{fig:tauslicesxy}) and the 3D data cubes are provided in fits format as supplementary materials. In each map, we plot the contrast fields $\Delta_{\tau}$ and $\Delta_{\log N}$ rather than the raw $\tau$ or $\log N_{\rm HI}$:
\begin{equation}\label{eq:tauconstrast}
  \Delta_{\tau}(\chi, x, y) \equiv \frac{\tau(\chi, x, y)}{\langle\tau\rangle} - 1,
\end{equation}
\begin{equation}\label{eq:Lyaconstrast}
  \Delta_{\log N}(\chi,x,y) \equiv \log_{10} N_{\rm HI}(\chi,x,y) - 
  \bigl\langle \log_{10} N_{\rm HI}\bigr\rangle,
\end{equation}
where $\langle\tau\rangle$ is the dimensionless volume-averaged optical depth over the entire cube, and $\bigl\langle \log_{10} N_{\rm HI}\bigr\rangle$ is the averaged logarithm of the \ion{H}{I} column density.

The geometry of the mock 3D gas distribution data cube is defined in comoving coordinates $(\chi, x, y)$, with a grid of size $N_\chi \times N_x \times N_y=2048\times512\times512$ spanning $z=1.0-1.4$ and an EDF-N-sized FOV, with transverse proper resolution $\sim 120\rm~pkpc$ and LOS resolution $\sim 140\rm~pkpc$. Such a data cube is not aiming at characterizing the large-scale gas distribution physically over a huge space volume, but rather used to justify the ability of the proposed absorption line tomography method in recovering larger-scale gas structures to a reasonable accuracy. As shown in Figs.~\ref{fig:NHIsliceszx} and \ref{fig:NHIslicesxy}, filamentary structures along the LOS and in the transverse directions are clearly visible from this mock data cube, with characteristic length scales defined by $L_{\rm corr}$. This data cube is therefore sufficient for the following analysis.

\subsection{Mock background galaxy catalog}\label{subsec:MockGalaxyCat}

In this subsection we construct a background galaxy catalog behind our primary foreground gas volume, tailored to the FOV of the EDF–N. Of course, galaxies at lower redshifts could also be used as background sources to probe lower–redshift gas structures via different lines, but here we focus on a configuration optimized for our fiducial foreground slab. The resulting catalog will later be used to simulate the CSST and JUST spectra and to recover the gas distribution built in \textsection\,\ref{subsec:MockGasCube}.

We adopt exactly the same evolving galaxy LF as in \textsection\,\ref{subsec:SrcDenDetLim} (Fig.~\ref{fig:NgalaxyAGN}), and apply it consistently here. The AGN contribution to the total number of background sources at the relatively bright detection limit adopted in this work is small (Table~\ref{tab:source_counts}; adopting $m_{\rm ref}^{\rm lim}\simeq23.5\rm~mag$ in a broad optical/NIR band, e.g., a $z$- or $J$-like band, approximately matched to the LF band), so we neglect AGN when building the mock catalog. We require background sources to lie well behind the foreground gas slab ($1.0 \le z \le 1.4$; \textsection\,\ref{subsec:MockGasCube}) and to have Ly$\alpha$ (for CSST) and \ion{Mg}{II} (for JUST) within the wavelength coverage. We therefore restrict the mock background galaxies to $z_{\rm g} \ge 1.5$; the exact high-$z$ cutoff is unimportant because the LF predicts a rapidly declining number of bright galaxies with increasing redshift (Fig.~\ref{fig:NgalaxyAGN}).

Operationally, we evaluate the LF on a regular grid in redshift and apparent magnitude over $1.0 \le z \le 4.0$ and $18 \le m_{\rm ref} \le m_{\rm ref}^{\rm lim}$, and compute the expected differential number counts per unit solid angle. Integrating over the EDF-N area ($20~\mathrm{deg}^2$) yields the expected number of galaxies in each $(z,m)$ bin. We then realize a discrete catalog whose total number of galaxies matches the LF prediction exactly by rounding the expected counts in each bin. Within each occupied $(z,m)$ bin we draw individual galaxy redshifts and magnitudes from uniform distributions across the bin boundaries, and we assign transverse $(x,y)$ positions by sampling uniformly over the projected footprint of the gas cube in comoving Mpc. The resulting table (in supplimentary material) includes 80,826 galaxies at $z=1.0-4.0$ and have $m_{\rm ref}=(18.0-23.5)\rm~mag$, of which, 5,394 are at $z\geq1.5$, forming the input catalog for the mock CSST and JUST spectra in \textsection\,\ref{subsec:CoupleGasGalaxy}. For each galaxy, the table stores its comoving coordinates $(x,y)$, redshift $z_{\rm g}$, and reference band magnitude $m_{\rm ref}$.

\begin{figure}
  \centering
  \includegraphics[width=0.49\textwidth]{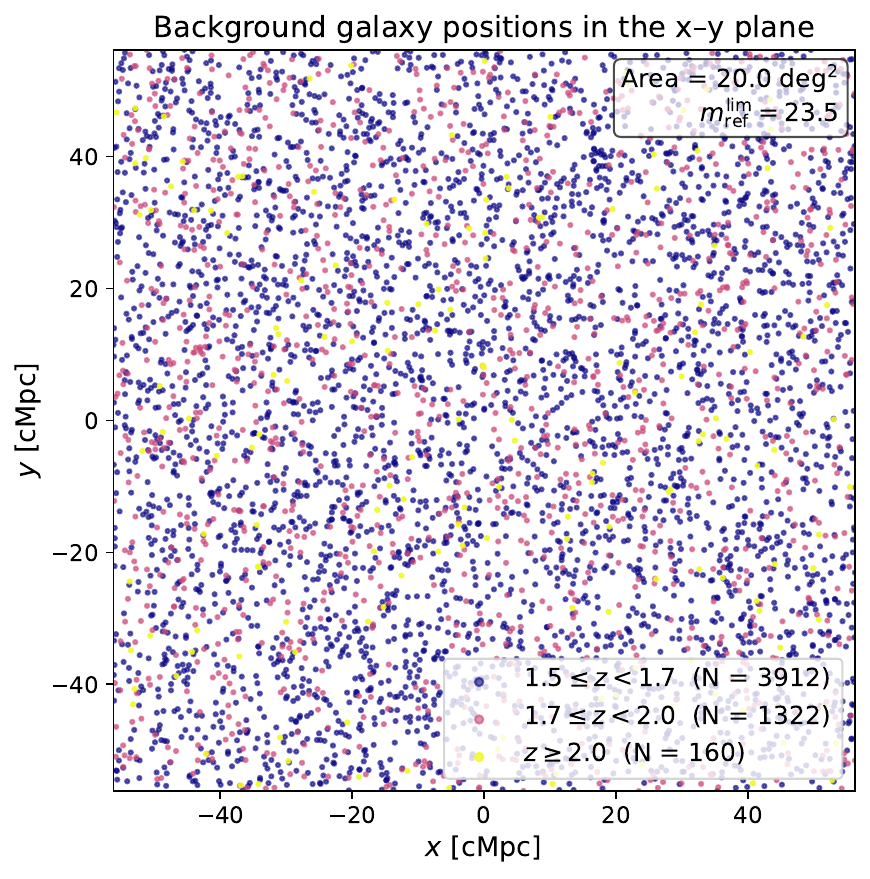}
  \caption{Spatial distribution of the mock background galaxy sample in the transverse comoving plane used for the EDF-N. Each point marks the $(x,y)$ position of a background galaxy brigther than a UV magnitude of 23.5 and with redshift $z\geq1.5$. Different colors correspond to distinct redshift intervals ($1.5 \leq z < 1.7$, $1.7 \leq z < 2.0$, and $z \geq 2.0$), and the legend reports the total number of galaxies in each bin.}\label{fig:bcksrcdistribution}
\end{figure}

Fig.~\ref{fig:bcksrcdistribution} shows the spatial distribution of the resulting background galaxies in the transverse comoving plane, color-coded by redshift in several bins (e.g. $1.5 \le z < 1.7$, $1.7 \le z < 2.0$, and $z \ge 2.0$). This figure can be compared to Fig.~\ref{fig:NHIslicesxy}, both have area comparable to EDF-N, in order to evaluate the sampling density of large-scale gas structures by background galaxies.

\subsection{Coupling the 3D gas distribution to background galaxies and mock spectra}
\label{subsec:CoupleGasGalaxy}

In this subsection we discuss the construction of realistic mock spectra for background galaxies at $z\gtrsim1.5$, observed with both the CSST slitless spectroscopic deep survey and the JUST multi-object spectrograph. We generate the model spectra using the \textsc{Bagpipes} code \citep{Carnall2018} plus some bright emission lines scaled to the SFR, following the same way as discussed in \textsection\,\ref{subsec:ExpPhotometry} and presented in Fig.~\ref{fig:sed_fit}. The detail parameters of the emission lines will not significantly affect our justification of the absorption line studies. To ensure consistent flux normalization across instruments, we compute an AB magnitude for each mock galaxy from the intrinsic SED using the CSST $z$-band filter transmission curve.

We emphasize that the CSST slitless spectra are not used here to directly detect spatially extended diffuse emission. Instead, they are employed to measure the Ly$\alpha$ forest absorption features imprinted on the spectra of discrete background galaxies. The spectral resolution and throughput of CSST are explicitly incorporated in the mock spectra generation described below. Since the analysis relies on 1D extracted spectra of individual background sources, rather than on spatially extended emission measurements, issues related to subtracting bright foreground sources along the spatial direction are not central to this approach.

We simulate the CSST slitless spectra in three bands (GU: $2,500-4,200~\text{\AA}$; GV: $4,000-6,500~\text{\AA}$; GU: $6,200-10,000~\text{\AA}$) using the throughput tables distributed with the CSST Emulator for Slitless Spectroscopy (CESS; \citealt{Wen2024}). We assume the depth from a typical deep survey with $8\times250\rm~s$ exposure, which reaches a 5$\sigma$ point-source depth of $\lesssim21.5$, 22.0, 22.0~mag in GU, GV, GI bands, respectively \citep{Gong2019}.

As JUST does not have a published simulator, we approximate the wavelength-dependent sensitivity using an empirical S/N template derived from an exposure time calculation: a 24~mag galaxy observed for $20\times1800\rm~s$ at $R\approx4000$. Such a template should be good enough for the purpose of this paper. We first restrict the intrinsic SED to the wavelength range of the template (approximately $3,500-10,000~\text{\AA}$) and convolve it with a Gaussian kernel to reach $R\approx4000$. The resulting spectrum $f_{\lambda,\mathrm{JUST}}(\lambda)$ is then sampled on the template wavelength grid. For a galaxy with AB magnitude $m_z$, and a total exposure time $t_{\mathrm{JUST}}$, we scale the S/N template as:
\begin{equation}
  \rm{SNR}_{\rm{JUST}}(\lambda) 
  = \rm{SNR}_{24}(\lambda)\,
    10^{-0.4\,(m_z - 24)}\,
    \sqrt{\frac{t_{\rm{JUST}}}{10~\rm{hr}}}.
\end{equation}
The per-pixel uncertainty is then computed analogously to the CSST case,
\begin{equation}
  \sigma_{f,\rm{JUST}}(\lambda) 
    = \frac{|f_{\lambda,\rm{JUST}}(\lambda)|}{\rm{SNR}_{\rm{JUST}}(\lambda)}.
\end{equation}

We next connect the 3D gas distribution from \textsection\,\ref{subsec:MockGasCube} to the mock background galaxy catalog from \textsection\,\ref{subsec:MockGalaxyCat} in order to generate realistic mock CSST and JUST spectra. From the mock galaxy catalog we select only ``background'' galaxies at $z_{\rm g} \ge 1.5$. Here we caution that in the galaxy catalog the SFR is scaled to the stellar mass $M_\star$ using the main sequence relation (e.g., \citealt{Speagle2014}), so the stellar mass is directly linked to the magnitude (converting from the rest-frame UV magntiude in the LF in \textsection\,\ref{subsec:SrcDenDetLim} to the CSST $z$-band magnitude at its redshift). The SFR factor $f_{\rm SFR}$, defined as $\rm{SFR} = f_{\rm SFR}\,\rm{SFR}_{\rm MS}$, is not really important is justifying the absorption line studies, as it mainly affect the emission line strength.

The 3D gas data cube provides the neutral hydrogen column density $N_{\rm HI}$ on a regular $(z, x_{\rm Mpc}, y_{\rm Mpc})$ grid. For a given background galaxy, we define a sightline by locating the nearest transverse grid cell at $(x_{\rm Mpc}, y_{\rm Mpc})$ and extracting the full LOS profile $N_{\rm HI}(z_i)$ along that column. We also compute the integrated \ion{H}{I} column density $N_{\rm HI,tot}$ over our desired absorber redshift range of $z=1.0-1.4$. Assuming a fixed ratio of $N_{\rm MgII}/N_{\rm HI}$ derived from a metallicity of $Z=0.3Z_\odot$, we further obtain the corresponding \ion{Mg}{II} column density profile and total \ion{Mg}{II} column density $N_{\rm MgII,tot}$ over $z=1.0-1.4$.

\begin{figure}
  \centering
  \includegraphics[width=0.49\textwidth]{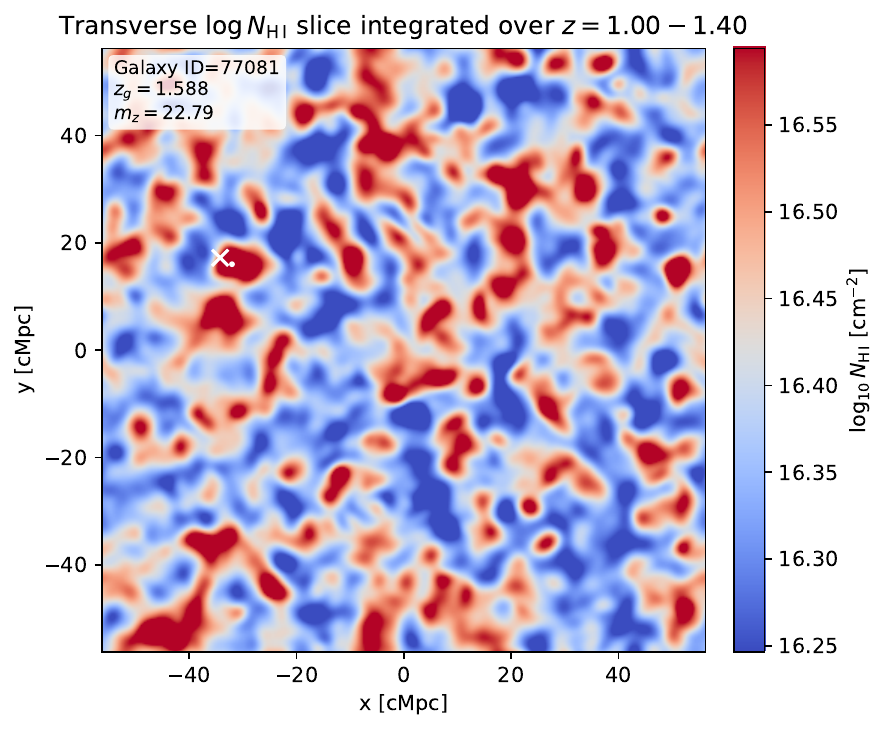}
  \caption{Transverse map of the neutral hydrogen column density integrated over $z=1.0-1.4$ from the 3D gas distribution cube (Figs.~\ref{fig:NHIsliceszx}, \ref{fig:NHIslicesxy}). The white cross marks the location of the background galaxy whose CSST and JUST spectra are presented in Fig.~\ref{fig:CSSTJUSTSpecRealCase}. The text box reports the galaxy ID in the galaxy catalog (\textsection\,\ref{subsec:MockGalaxyCat}), redshift, and CSST $z$-band magnitude.}\label{fig:transverseNHIz1014}
\end{figure}

We then imprint the LOS absorption due to Ly$\alpha$ and the \ion{Mg}{II} doublet using the \ion{H}{I} and \ion{Mg}{II} column density profiles. For each transition with rest-frame wavelength $\lambda_0$ and oscillator strength $f_{\rm osc}$, we adopt a Doppler profile with a width of $b_{\rm Ly\alpha}=b_{\rm Mg\,II}=27\rm~km~s^{-1}$, matching the velocity resolution of our data cube (\textsection\,\ref{subsec:MockGasCube}). This is an oversimplied assumption, as most of the IGM \ion{Mg}{II} absorbers may be indeed narrower. We then compute the optical depth as:
\begin{equation}
  \tau(\lambda) = \sum_i \tau_{0,i} \exp\left[-\left(\frac{v(\lambda, z_i)}{b}\right)^2\right],
\end{equation}
where $\tau_{0,i} \propto \lambda_0 f_{\rm osc} N_i / b$, $N_i$ is the column density in the $i$th redshift bin. In the absorber redshift interval $z=1.0-1.4$ the typical integrated columns densities in our cube are $N_{\rm HI}\sim 10^{16.5-17}\rm~cm^{-2}$ (see Fig.~\ref{fig:transverseNHIz1014}) and $N_{\rm Mg\,II}\sim 10^{12-12.5}\rm~cm^{-2}$.

\begin{figure*}
  \centering
  \includegraphics[width=0.49\textwidth]{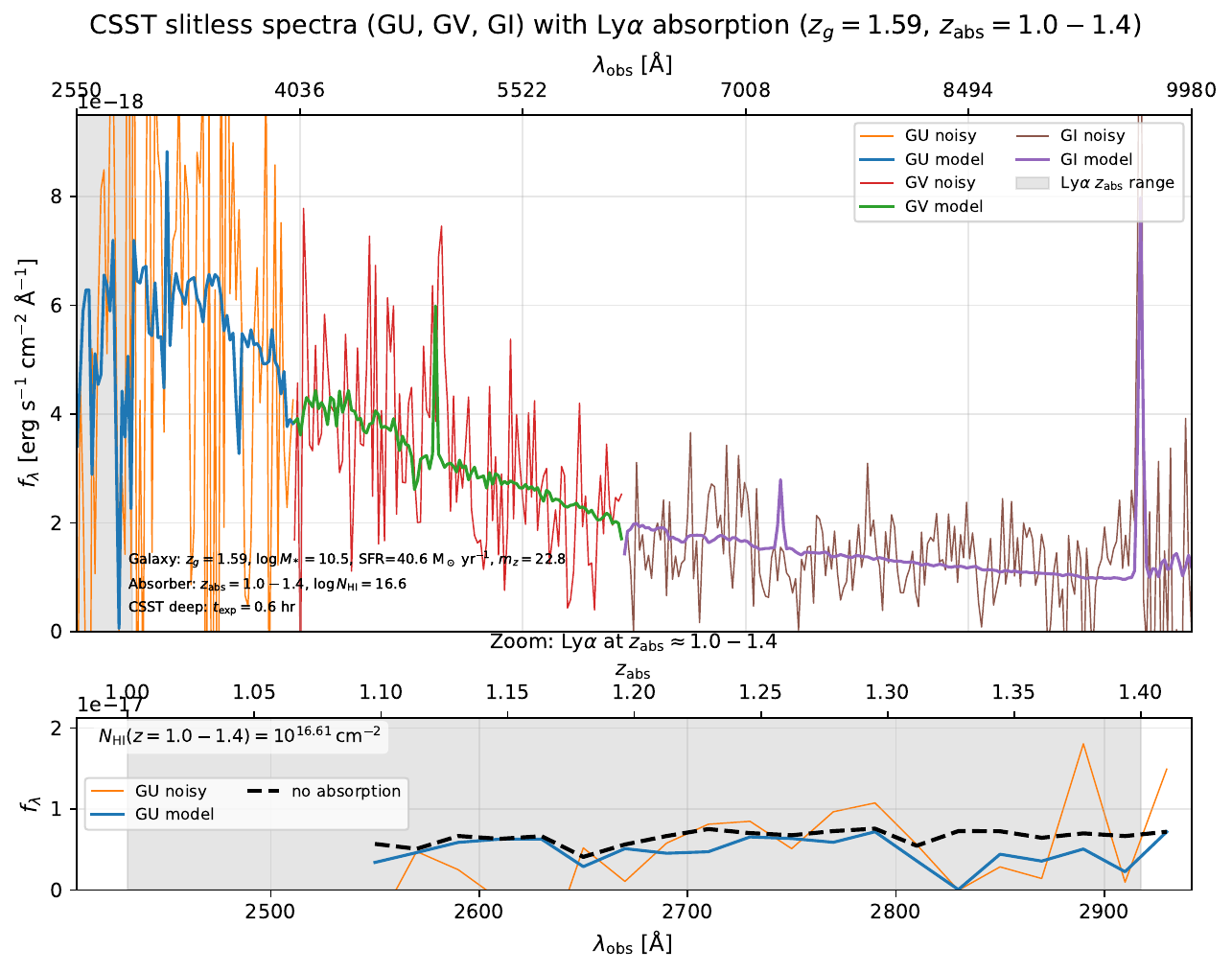}
  \includegraphics[width=0.49\textwidth]{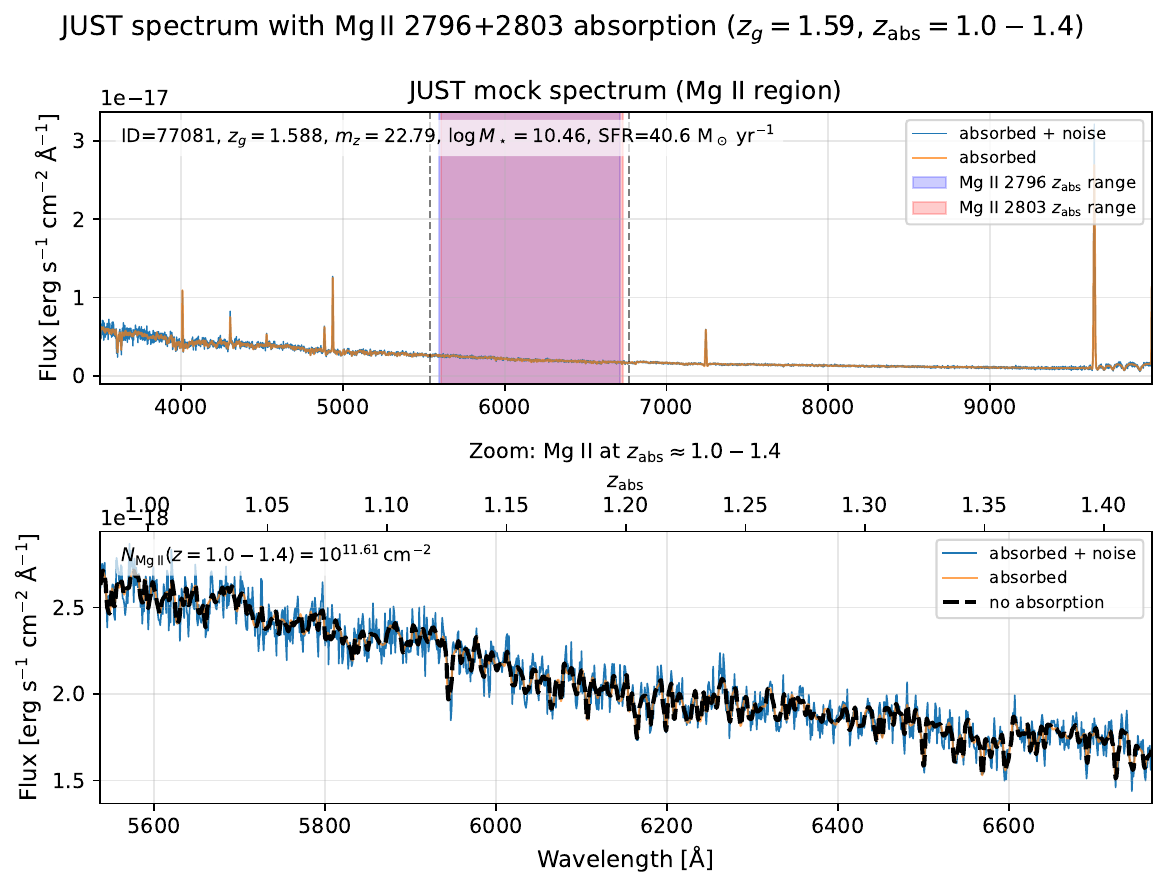}
  \caption{Mock CSST (left) and JUST (right) spectra for a representative background galaxy at $z_{\rm g}\simeq1.59$ (Fig.~\ref{fig:transverseNHIz1014}). \emph{Top panels:} full-band CSST slitless spectra in the GU, GV, and GI bands (model and noisy realizations) and the JUST slit spectrum, with the wavelength intervals corresponding to Ly$\alpha$ and the \ion{Mg}{II} doublet in the redshift range $z=1.0-1.4$ highlighted. \emph{Bottom panels:} zoom-in views around Ly$\alpha$ (CSST) and the \ion{Mg}{II} $\lambda2796$ and $\lambda2803$ doublet (JUST) approximately in the redshift range of $z\sim1.0-1.4$. The two shaded regions for the \ion{Mg}{II} doublet largely overlap at this resolution, but are shown separately to indicate the wavelength coverage of the individual transitions. In each zoom-in panel we show the absorbed model spectrum, a noisy realization at the adopted exposure times (CSST deep survey with $8\times250\rm~s$ and a 10~hr JUST exposure), and a thick dashed curve corresponding to the absorption-free galaxy spectrum at the same instrumental resolution. Given the typical total integrated column densities over $z=1.0-1.4$ predicted by our gas cube ($N_{\rm HI, tot}\lesssim 10^{17}\,\mathrm{cm^{-2}}$, $N_{\rm Mg\,II, tot}\lesssim 10^{12.5}\,\mathrm{cm^{-2}}$), the Ly$\alpha$ forest is detectable in the CSST spectra, although with limited velocity resolution, whereas individual \ion{Mg}{II} absorbers are in general too weak to be detected in single JUST spectra at the assumed depth.}\label{fig:CSSTJUSTSpecRealCase}
\end{figure*}

Using the Ly$\alpha$- or \ion{Mg}{II}-absorbed model spectra, we simulate CSST slitless or JUST fiber spectra following the procedure described above. For comparison we also generate spectra at the same resolution from the absorption-free model. The resulting full-band and zoomed views demonstrate that Ly$\alpha$ absorption in the $z=1.0-1.4$ slab is detectable and marginally resolvable in the CSST spectra, albeit with limited velocity resolution (Fig.~\ref{fig:CSSTJUSTSpecRealCase}). However, given the relatively modest \ion{Mg}{II} column densities predicted on the scales resolved by our gas cube, individual \ion{Mg}{II} systems are generally not detectable in single JUST spectra at these depths (Fig.~\ref{fig:CSSTJUSTSpecRealCase}). Nevertheless, compact, higher column density systems below our grid resolution could still produce strong \ion{Mg}{II} features (e.g., \citealt{LiJ2025}), and, more generally, \ion{Mg}{II} and other even weaker metal lines may still be detectable statistically by stacking many sightlines aligned in the rest frame. This will be explored in later sections (\textsection\,\ref{subsec:MgIIstacking} and \ref{subsec:MgII2PCF}).

For all the background galaxies (we adopt $z\geq1.5$), we apply the above approach to generate its CSST and JUST spectra, adding Ly$\alpha$ and \ion{Mg}{II} absorption lines, respectively. The full spectral catalog of 5,394 background galaxies (Fig.~\ref{fig:bcksrcdistribution}) is provided as supplementary materials and introduced in detail in the Appendix \textsection\,\ref{Appsec:SpecCatBckGal}. This catalog will be used in the follow-up statistical analyses.

\subsection{Reconstructing the \ion{H}{I} distribution with the apparent optical depth method}\label{subsec:AODCSSTreconstruction}

A natural way to connect our mock galaxy CSST spectra to the underlying 3D \ion{H}{I} distribution is to work directly with the apparent optical depth (AOD) of the Ly$\alpha$ forest. We follow the AOD formalism of \citet{Savage1991}, which has become a standard tool for converting normalized absorption profiles into column densities without explicit profile decomposition, and for quantifying the impact of noise and unresolved structure (e.g., \citealt{Fox2005}). The AOD method is optimized for our low-resolution CSST spectra, where the decomposition of different absorption components is difficult, while the continuum can be well characterized thanks to the wide wavelength coverage.

Rather than re-fitting the continuum, we exploit the galaxy parameters stored in the catalog to construct an essentially noise-free model continuum, following the same approach as discussed above. This approach is just for a simple justification in this paper, while in the real case, the continuum determination strongly depends on the modeling procedure of the CSST spectra of background galaxies. We then propagate the model SED through the CSST instrument response to obtain the expected continuum $F_{\lambda,{\rm cont}}(\lambda)$. We focus on the Ly$\alpha$ forest region, which generally falls in the GU band for the redshift range of interest. For each galaxy we take the Ly$\alpha$-absorbed mock spectrum $F_\lambda(\lambda)$ from the galaxy spectral catalog constructed in \textsection\,\ref{subsec:CoupleGasGalaxy} and compute the AOD on a pixel-by-pixel basis,
\begin{equation}
    \tau_{\rm a}(\lambda) \;=\; -\ln \left[ \frac{F_\lambda(\lambda)}{F_{\lambda,{\rm cont}}(\lambda)} \right].
\end{equation}
The $\tau_{\rm a}$ profiles for all galaxies are interpolated onto a common absorber redshift grid with $z_{\rm abs} \approx 1.1-1.4$ and $N_z = 256$ bins. This yields a two-dimensional (2D) AOD cube, $\tau_{\rm a}^{\rm LOS}(i, z_j)$, where $i$ indexes the background galaxies (or sightlines) and $z_j$ labels the absorber redshift bins.

To convert $\tau_{\rm a}$ into (a lower limit of) $N_{\rm HI}$ we use the standard AOD relation for a single transition \citep[][their Eq.~4]{Savage1991},
\begin{equation}
    N_{\rm a} \;=\; \frac{m_e c}{\pi e^2} \;\frac{1}{f \lambda_0}
    \int \tau_{\rm a}(v)\,{\rm d}v,
\end{equation}
where $f$ and $\lambda_0$ are the oscillator strength and rest-frame wavelength of Ly$\alpha$, and $v$ is velocity in the absorber rest frame. In our discrete implementation we treat each absorber redshift bin as a small segment in comoving path length $\Delta \chi$ and convert the $\tau_{\rm a}(z)$ field into an estimate of $N_{\rm HI}$ per unit comoving distance,
\begin{equation}
    \frac{{\rm d}N_{\rm HI}}{{\rm d}\chi}(z_j) \;\approx\; 
    {\cal C}_{\rm Ly\alpha}\,\tau_{\rm a}(z_j)\,
    \frac{H(z_j)}{1+z_j},
\end{equation}
where $H(z)$ is the Hubble constant and ${\cal C}_{\rm Ly\alpha}$ is the Ly$\alpha$-specific normalization corresponding to the above AOD expression. For each LOS we then obtain a one-dimensional (1D) profile $N_{\rm HI}^{\rm LOS}(\chi)$ by multiplying ${\rm d}N_{\rm HI}/{\rm d}\chi$ by the comoving cell width $\Delta\chi$ at each redshift.

In order to compare directly with the 3D gas distribution data cube, we reconstruct an ``AOD cube'' $N_{\rm HI}^{\rm AOD}(x,y,\chi)$ on exactly the same $(2048, 512, 512)$ grid. Transversely, we adopt a nearest-neighbour or Voronoi-like assignment (e.g., \citealt{Cappellari2003,LiJ2015}). For every $(x,y)$ cell center of the simulation grid we find the nearest galaxy LOS in the plane of the sky, based on the comoving coordinates $(x_i,y_i)$ of the background galaxies. All transverse positions whose nearest galaxy is sightline $i$ are assigned that LOS's \ion{H}{I} column density profile $N_{\rm HI}^{\rm LOS}(\chi)$. Along the LOS direction we linearly interpolate $N_{\rm HI}^{\rm LOS}(\chi)$ from the coarse AOD redshift grid ($N_z = 256$ in $z\approx1.0-1.4$) onto the finer simulation grid ($N_z=2048$ between $z=1.0-1.4$; simply assuming $N_{\rm HI}=0$ at $z=1.0-1.1$), for the convenience of comparison with the 3D gas distribution data cube constructed in \textsection\,\ref{subsec:MockGasCube}.

To validate the reconstruction we compare transverse maps of the AOD cube and the true simulation gas distribution cube from \textsection\,\ref{subsec:MockGasCube} after integrating the \ion{H}{I} field over the main redshift range of interest (here we adopt $z=1.15-1.4$). We then examine both $\log_{10} N_{\rm HI}^{\rm int}(x,y)$ and the fractional residual map defined as: 
\begin{equation}\label{eq:AODNHIresidual}
    \Delta(x,y) \;=\;
    \frac{N_{\rm HI,\,AOD}^{\rm int}(x,y) -
          N_{\rm HI,\,true}^{\rm int}(x,y)}
         {N_{\rm HI,\,true}^{\rm int}(x,y)}.
\end{equation}

\begin{figure*}
  \centering
  \includegraphics[width=0.49\textwidth]{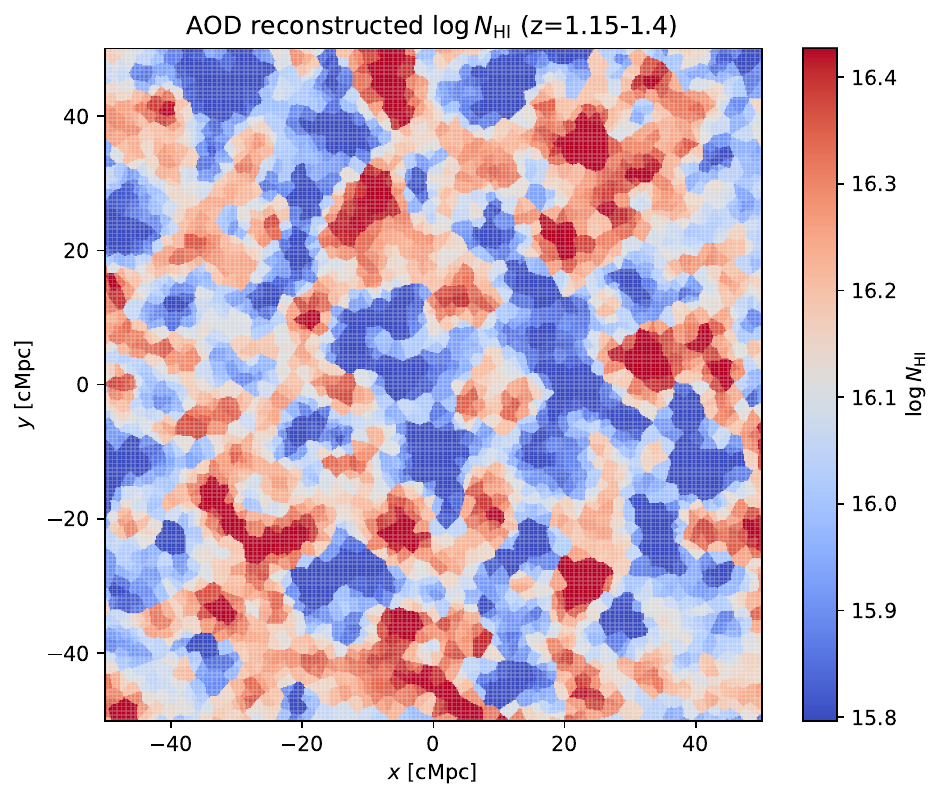}
  \includegraphics[width=0.49\textwidth]{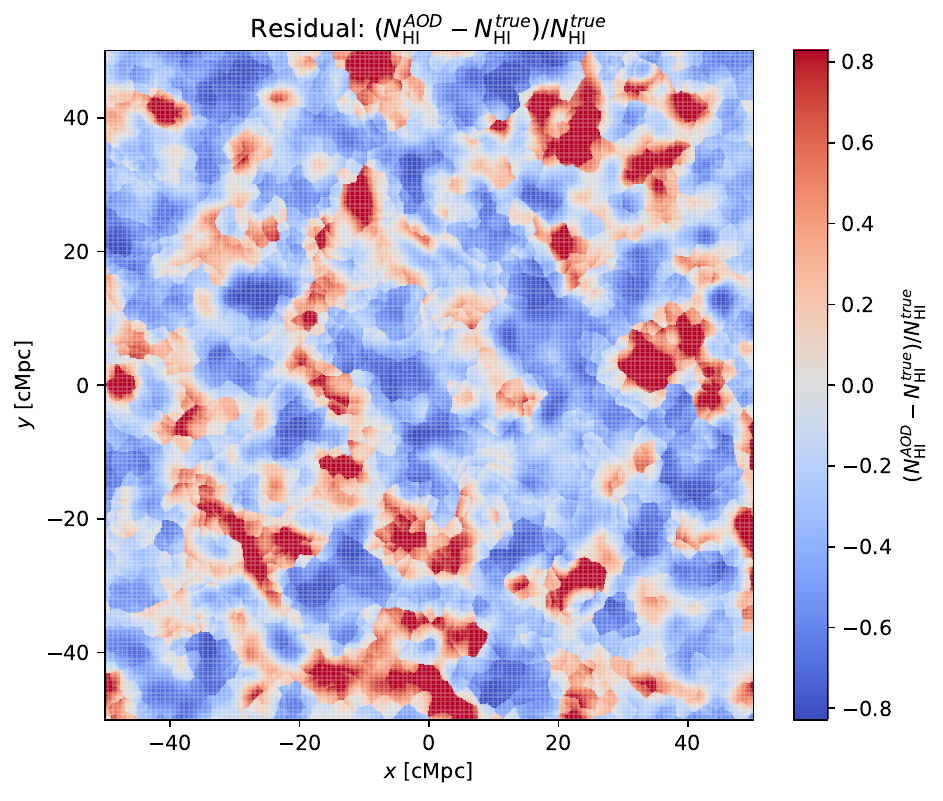}
  \caption{\emph{Left}: Transverse map of the AOD-reconstructed \ion{H}{I} column density from the mock CSST spectra of $z\geq1.5$ background galaxies (Fig.~\ref{fig:bcksrcdistribution}), integrated over $z=1.15-1.4$. \emph{Right}: Fractional residual map (Eq.~\ref{eq:AODNHIresidual}) comparing the AOD reconstruction to the true simulation cube (similar as Fig.~\ref{fig:transverseNHIz1014}).}\label{fig:AODRecon_xy}
\end{figure*}

Fig.~\ref{fig:AODRecon_xy} shows that, despite the sparse and slightly anisotropic sampling by background galaxies (Fig.~\ref{fig:bcksrcdistribution}), the AOD reconstruction captures the large-scale structure of the \ion{H}{I} field reasonably well. Filaments and voids are reproduced on scales comparable to the mean LOS separation ($\sim1\rm~cMpc$), while small–scale features below the transverse sampling scale are naturally smoothed out by the Voronoi-like interpolation. Further biases and scatters may come from the spatial sampling: (i) regions far from any LOS, where the Voronoi assignment extrapolates a single LOS profile over a large area, and (ii) high-$N_{\rm HI}$ peaks where Ly$\alpha$ saturation and strong peculiar velocities challenge the simple AOD conversion. These limitations are well understood from previous applications of the AOD technique to noisy, blended absorption profiles (e.g., \citealt{Fox2005}) and will be important when designing LOS sampling strategies for real surveys. The AOD reconstructed gas distribution is also provided as supplementary data as detailed in \textsection\,\ref{Appsec:AODGasCubeData}.

To further assess the reliability of the AOD reconstruction, we quantitatively compare the LOS integrated \ion{H}{I} column densities recovered from the mock CSST spectra, $N_{\rm HI,AOD}$, to the true columns directly summed along the corresponding LOS through the simulation cube, $N_{\rm HI,true}$. For our fiducial setup we obtain a mean ratio $\langle N_{\rm{HI,AOD}}/N_{\rm{HI,true}}\rangle \simeq 0.88$, with low-$N_{\rm HI}$ sightlines typically biased low and high-$N_{\rm HI}$ sightlines often slightly above unity (Fig.~\ref{fig:AODRecon_xy}). This behavior is expected once both finite spectral resolution and realistic noise are taken into account.

In the idealized, noise-free limit, the AOD method is known to recover accurate column densities for resolved, unsaturated lines, but to underestimate $N_{\rm HI}$ when strong, narrow components are spectrally unresolved and thus appear as shallower, broader features in the convolved spectrum (e.g., \citealt{Savage1991,Jenkins1996}). This ``unresolved saturation'' effect preferentially impacts high-$N_{\rm HI}$ systems and leads to $N_{\rm HI}^{\rm AOD}$ being a lower limit to the true column density, a trend that has also been highlighted in recent comparisons between empirical AOD analyses and more physical line-transfer models (e.g., \citealt{Huberty2024}). At the same time, the non-linear mapping from flux to optical depth implies that, in the presence of noise, weak lines near the detection limit are particularly vulnerable: random noise fluctuations and any masking or clipping of negative AOD can cause the integrated $N_{\rm HI}$ of low-$N_{\rm HI}$ sightlines to be systematically underestimated, while at very low S/N the AOD method can even overestimate columns for individual weak features \citep{Fox2005}. The fact that our global mean $N_{\rm HI,AOD}/N_{\rm HI,true}$ is slightly below unity therefore indicates that, in our mock CSST spectra, unresolved saturation and the loss of weak absorption in noisy data both contribute to a net modest underestimation of the true \ion{H}{I} columns, as well as the positive spatial correlation between the residual (Eq.~\ref{eq:AODNHIresidual}) and the $N_{\rm HI}$.

An alternative approach to studying the Ly$\alpha$ forest with CSST is through auto- and cross-correlation analyses of Ly$\alpha$ absorption measured along background quasar and galaxy sightlines (e.g., \citealt{Tan2025}). In contrast, our work explores a complementary regime by leveraging the much higher surface density of background galaxies selected via combined Euclid, CSST, and JPCam photometry. This strategy enables direct absorption line stacking and spatially resolved tomographic reconstruction of the IGM, providing a more localized and structure-oriented view of large-scale gas structures at $z\sim1-2$.

\subsection{Constraints on \ion{Mg}{II} absorption from stacked JUST spectra} \label{subsec:MgIIstacking}

The strongest metal absorption lines covered by the higher-resolution JUST spectra are the \ion{Mg}{II} doublet. We showed above (Fig.~\ref{fig:CSSTJUSTSpecRealCase}) that these IGM \ion{Mg}{II} absorbers are generally not individually detectable at the current survey depth. However, with $\gtrsim5,000$ background galaxies usable for absorption line studies, we may still find some statistical way to constrain the IGM metallicity based on the JUST spectra. The first method we will be exploring is the direct stacking of weak \ion{Mg}{II} signals. To assess the feasibility of detecting weak IGM \ion{Mg}{II} absorption via this method, we stack the mock background galaxy spectra in the rest frame of absorbers identified along the CSST GU Ly$\alpha$ forest sightlines. In all cases we work with normalized JUST spectra, $F(\lambda) / F_{\rm cont}(\lambda)$, and average the transmitted flux over many LOS. We contrast two limiting strategies for defining the absorber redshift:
(i) an ``ideal'' selection in which the \ion{Mg}{II} absorber positions are taken directly from the noiseless input 3D gas distribution data cube (\textsection\,\ref{subsec:MockGasCube}), and
(ii) a CSST-based selection in which absorbers are identified in the low-resolution Ly$\alpha$ forest via an AOD threshold $\tau_{\rm Ly\alpha,min}$ and their redshifts are used to stack the JUST \ion{Mg}{II} spectra.

\begin{figure}
\centering
\includegraphics[width=0.48\textwidth]{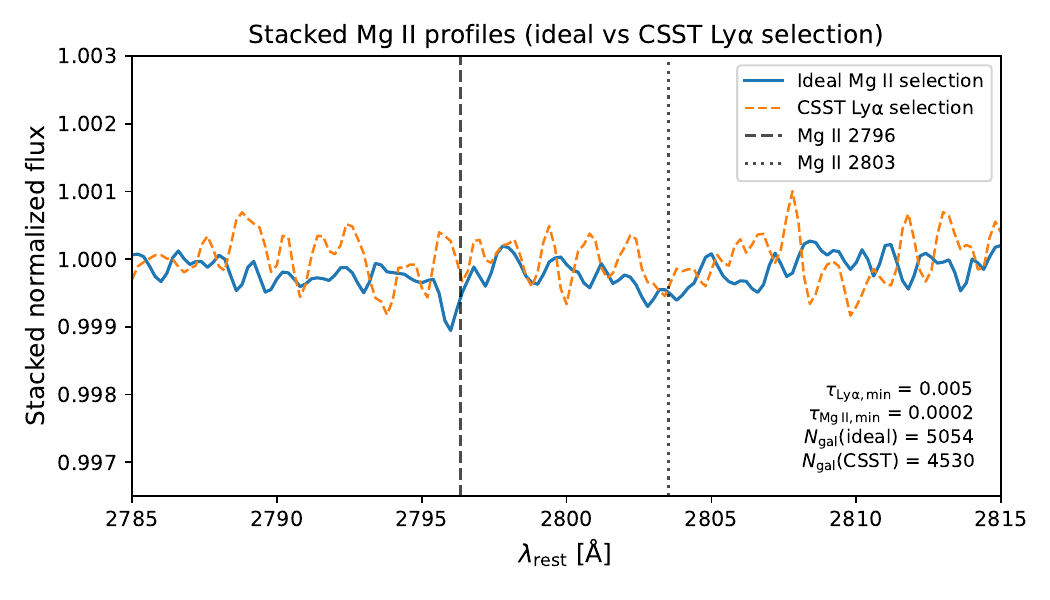}
\caption{Stacked \ion{Mg}{II} absorption profiles in the JUST spectra for the ``ideal'' (blue, solid) and CSST Ly$\alpha$ selected (orange, dashed) absorber samples, plotted in the absorber rest frame. The vertical dashed and dotted lines mark the \ion{Mg}{II} $\lambda2796~\text{\AA}$ and $\lambda2803~\text{\AA}$ transitions, respectively. In this example we adopt a Ly$\alpha$ AOD threshold of $\tau_{\mathrm{Ly\alpha},\min}=0.005$, which corresponds to a \ion{Mg}{II} threshold of $\tau_{\mathrm{Mg\,II},\min}\approx 2\times10^{-4}$ for the assumed metallicity and ionization state; the number of contributing galaxies in each stack is indicated in the legend.}\label{fig:MgIIStackIdealCSST}
\end{figure}

We show an example of the stacked \ion{Mg}{II} 2796/2803~\text{\AA} profiles in Fig.~\ref{fig:MgIIStackIdealCSST} for a representative Ly$\alpha$ threshold $\tau_{\rm Ly\alpha,min}\simeq0.005$. In this experiment the corresponding \ion{Mg}{II} threshold in the ideal selection is obtained by assuming a metallicity $Z\simeq0.3~Z_\odot$ and a characteristic ionization state, such that $\tau_{\rm MgII,min}$ scales with $\tau_{\rm Ly\alpha,min}$ according to: 
\begin{equation}
\frac{\tau_{\rm MgII}}{\tau_{\rm Ly\alpha}}
\;\approx\;
\frac{N_{\rm MgII}}{N_{\rm HI}}\,
\frac{f_{\rm MgII}\lambda_{\rm MgII}}{f_{\rm Ly\alpha}\lambda_{\rm Ly\alpha}}.
\end{equation}
As shown in Fig.~\ref{fig:MgIIStackIdealCSST}, the ideal stack (using the true \ion{Mg}{II} absorber redshifts) shows a clear detection of the stronger \ion{Mg}{II} 2796~\text{\AA} line (the weaker 2803~\text{\AA} line is probably not detected), whereas the stack constructed using CSST Ly$\alpha$ absorber redshifts remains nearly featureless. Therefore, the gain in number of spectra does not compensate for the loss of coherence caused by redshift uncertainties and line blending in the Ly$\alpha$ forest.

\begin{figure}
\centering
\includegraphics[width=0.48\textwidth]{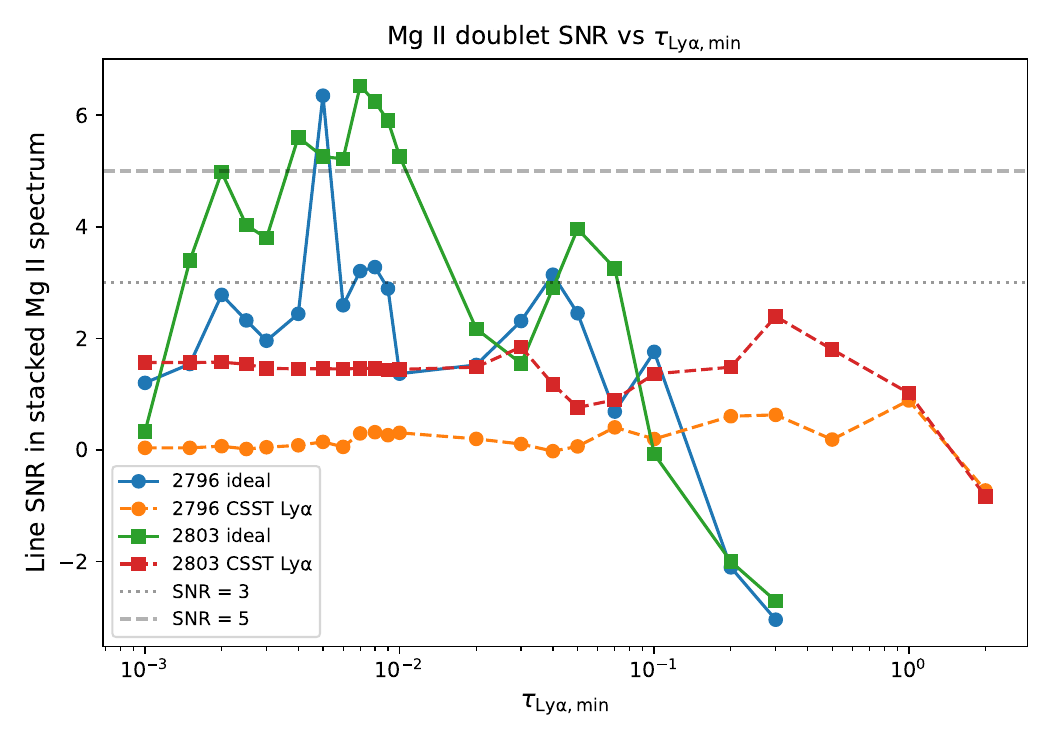}
\caption{Detection significance of the \ion{Mg}{II} doublet in stacked JUST spectra as a function of the adopted minimum Ly$\alpha$ apparent optical-depth $\tau_{\mathrm{Ly\alpha},\min}$. The blue circles and green squares show the S/N of the \ion{Mg}{II} $\lambda2796~\text{\AA}$ and $\lambda2803~\text{\AA}$ lines, respectively, for the ideal selection in which absorber redshifts are taken directly from the simulated 3D gas distribution data cube in \textsection\,\ref{subsec:MockGasCube}. The orange circles and red squares show the corresponding S/N values when absorbers are selected from the CSST Ly$\alpha$ forest using the same $\tau_{\mathrm{Ly\alpha},\min}$, and the stacked \ion{Mg}{II} spectrum is shifted by the CSST-derived Ly$\alpha$ absorber redshifts. Horizontal grey lines indicate S/N$=3$ (dotted) and S/N$=5$ (dashed).}\label{fig:MgIISNRvsTaumin}
\end{figure}

We quantify this behavior by measuring the detection significance of the \ion{Mg}{II} doublet as a function of the adopted minimum Ly$\alpha$ optical depth $\tau_{\rm Ly\alpha,min}$. For each value of $\tau_{\rm Ly\alpha,min}$ we (i) convert it to the corresponding $\tau_{\rm MgII,min}$ in the ideal case, (ii) stack the noisy JUST spectra in the absorber rest frame, and (iii) measure the equivalent width and S/N of the \ion{Mg}{II} 2796 and 2803~\text{\AA} lines using sidebands to estimate the local continuum and noise. The resulting S/N curves are shown in Fig.~\ref{fig:MgIISNRvsTaumin}. The ``ideal'' selection reaches a maximum S/N of $\sim 6-7$ for both the 2796~\text{\AA} and 2803~\text{\AA} line at intermediate thresholds ($\tau_{\rm Ly\alpha,min}\sim {\rm few}\times10^{-3}$), where the stacked sample balances column density against the number of contributing sightlines. In contrast, the CSST-based stacks never approach a statistically significant detection: the S/N of the both the 2796~\text{\AA} and 2803~\text{\AA} lines are consistent with pure noise for all thresholds.

These tests demonstrate that using the CSST Ly$\alpha$ forest alone as a redshift indicator for individual gas absorbers is not accurate enough to recover the weak IGM \ion{Mg}{II} signal in stacked JUST spectra. The limited spectral resolution and S/N of the CSST GU band smear the Ly$\alpha$ line profiles and introduce redshift errors that are large compared to the intrinsic width of the \ion{Mg}{II} doublet in the higher-resolution JUST spectra. When the stacked spectra are shifted according to these uncertain Ly$\alpha$ redshifts, the \ion{Mg}{II} absorption is effectively washed out, even though the total number of background galaxies ($\gtrsim 5,000$) would in principle be sufficient to detect the signal if the absorber redshifts were known precisely.

On the other hand, the ideal-selection experiment shows that if we can locate the absorbers in redshift space with substantially higher precision --- for example by using foreground galaxies, galaxy groups, or other large-scale structures as redshift anchors, or by exploiting higher-resolution Ly$\alpha$ (or other strong UV absorption lines such as \ion{C}{IV}; e.g., \citealt{LiJ2025,Yu2025}) forest spectroscopy of the same systems --- the \ion{Mg}{II} doublet becomes detectable at acceptable S/N in stacked JUST spectra. Furthermore, our mock gas data cube does not include compact and denser CGM absorbers, which could produce strong enough \ion{Mg}{II} absorption signals even detectable around individual galaxies (e.g., \citealt{Tie2024,LiJ2025}).

\subsection{Measuring diffuse IGM \ion{Mg}{II} with the two-point correlation function}\label{subsec:MgII2PCF}

An alternative route to measuring the diffuse IGM metallicity with the JUST spectra is to treat the \ion{Mg}{II} absorption as a continuous ``forest'' and to infer its amplitude statistically from the two-point correlation function (2PCF) of the transmitted flux. This method was developed by \citet{Hennawi2021}, who showed that the auto-correlation of the \ion{Mg}{II} optical depth fluctuations exhibits a narrow peak at the doublet separation ($\Delta v \simeq 768\rm~km~s^{-1}$) whose amplitude encodes the large-scale mean \ion{Mg}{II} opacity and hence the volume-averaged metallicity of the IGM. A key advantage of this technique over traditional line identification or stacking methods is that it uses \emph{all} pixels in the forest, is insensitive to line blends or incompleteness in discrete absorber catalogs, and does not require associating absorbers with individual galaxies (for an accurate redshift assignment, as discussed above in \textsection\,\ref{subsec:MgIIstacking}). The method was subsequently tested observationally by \citet{Tie2024} (and theoretically for \ion{C}{IV} by \citealt{Tie2022}), who measured a strong small-scale \ion{Mg}{II} 2PCF signal dominated by discrete CGM absorbers and demonstrated that the correlation becomes consistent with noise once these strong systems are masked.

\begin{figure*}
    \centering
    \includegraphics[width=1.0\linewidth]{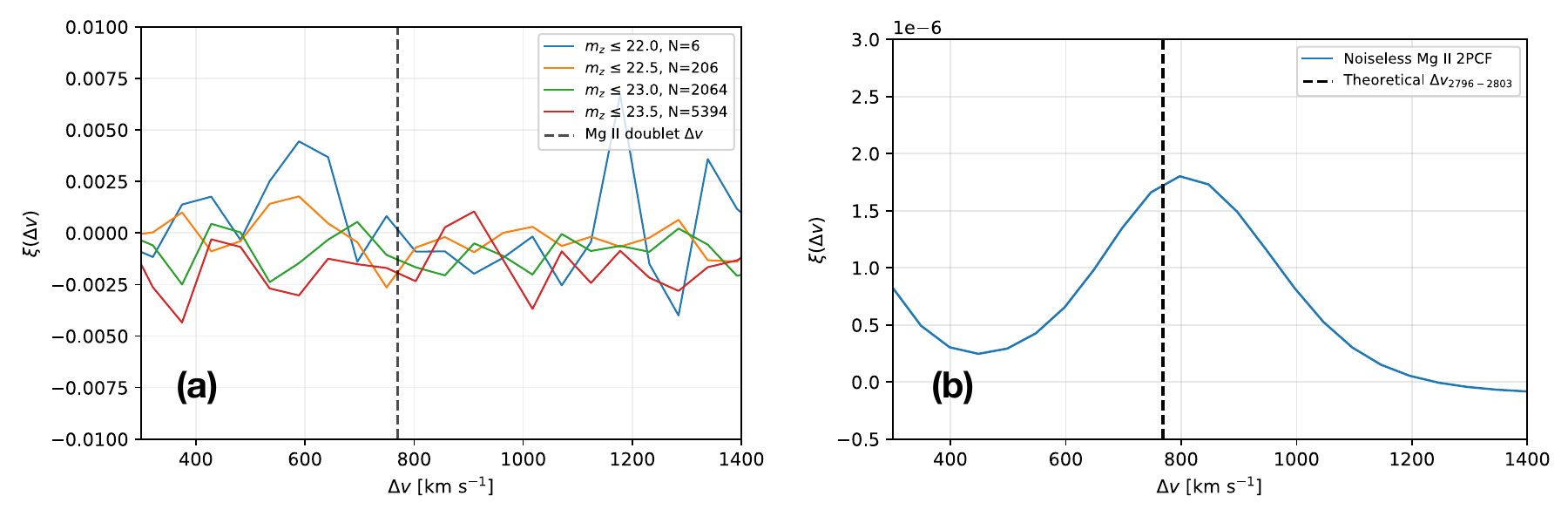}
    \caption{\ion{Mg}{II} two-point correlation function (2PCF) in the mock JUST survey. \textbf{Left:} 1D auto-correlation $\xi(\Delta v)$ of the \ion{Mg}{II} flux contrast field measured from the noisy mock JUST spectra, for different CSST $z$-band magnitude limits ($m_z \le 22.0$, 22.5, 23.0, and 23.5, as indicated in the legend). All curves are consistent with zero within their rms fluctuations ($\sigma_\xi \sim 10^{-3}$ per bin), and no significant peak is visible at the \ion{Mg}{II} doublet separation $\Delta v \simeq 768~\mathrm{km\,s^{-1}}$ (marked with the vertical dashed line).
    \textbf{Right:} Corresponding \ion{Mg}{II} 2PCF measured from the underlying noiseless model spectra, which exhibit a clear but very small peak at the doublet separation with amplitude $\xi_{\rm peak} \approx 2\times10^{-6}$.}\label{fig:MgII2PCF}
\end{figure*}

We follow this approach by constructing a \ion{Mg}{II} ``forest'' in our mock JUST spectra and computing its 1D auto-correlation along the LOS. For each background galaxy, we select the wavelength range corresponding to \ion{Mg}{II} $\lambda\lambda2796,2803~\text{\AA}$ absorption over the redshift interval of interest, resample the spectra onto a common logarithmic wavelength (or velocity) grid, and define the flux contrast as $\delta_F = F / \langle F\rangle - 1$. We then compute the 1D 2PCF $\xi(\Delta v)$ by averaging $\delta_F(v)\,\delta_F(v + \Delta v)$ over all pixels and all sightlines. Fig.~\ref{fig:MgII2PCF}a shows the resulting \ion{Mg}{II} 2PCF computed from the mock \emph{noisy} JUST spectra for different $z$-band magnitude limits ($m_z \le 22.0$, 22.5, 23.0, and 23.5, corresponding to increasingly larger samples and lower per-object S/N). All curves are consistent with zero within their noise, with typical rms fluctuations of order $\sigma_\xi \sim 10^{-3}$ per velocity bin and no visible peak at the expected doublet separation of $\Delta v \simeq 768\rm~km~s^{-1}$. In contrast, Fig.~\ref{fig:MgII2PCF}b shows the 2PCF measured from the underlying \emph{noiseless} model spectra, which exhibit a clear but extremely small peak at the \ion{Mg}{II} doublet separation with an amplitude of order $\xi_{\rm peak} \approx 2\times10^{-6}$.

The comparison in Fig.~\ref{fig:MgII2PCF} immediately explains why the \ion{Mg}{II} 2PCF signal is undetected in our realistic mock survey. For our fiducial model of diffuse IGM enrichment, the expected \ion{Mg}{II} doublet peak is roughly three orders of magnitude smaller than the noise level of the 2PCF measured from the noisy JUST spectra. In our EDF-N mock, the deepest sample ($m_z \le 23.5$) contains $N_0 = 5394$ background galaxies and yields a formal detection significance of only ${\rm SNR}_0 \simeq \xi_{\rm peak}/\sigma_\xi \sim 2\times10^{-3}$. Because the uncertainty of the 1D correlation function scales approximately as $\sigma_\xi \propto \rm{SNR}_{\rm spec}^{-2} N_{\rm LOS}^{-1/2}$, achieving a $5\sigma$ detection of the IGM-only \ion{Mg}{II} 2PCF at fixed survey depth would require of order $N_{\rm LOS} \sim 10^{9-10}$ bright background galaxies --- many orders of magnitude more than available in any foreseeable survey. Alternatively, keeping the number of sightlines fixed at $N_0$ and increasing only the exposure time, the 2PCF significance in the photon-noise-dominated regime scales as ${\rm SNR}_{\rm 2PCF} \propto t$, so a factor of $\sim10^3$ longer exposure (i.e., $\sim 10^4\rm~h$ per field) would be required to reach ${\rm SNR}\sim5$. Even combining brighter targets with a very wide-area survey (e.g., $A\sim10^4\rm~deg^2$, about 500 times the EDF-N) only increases ${\rm SNR}_{\rm 2PCF}$ by a factor of $\sim20-30$ relative to our fiducial setup, still far short of a statistically significant detection of the IGM-only signal.

These rough scaling arguments imply that the diffuse, smoothly enriched IGM \ion{Mg}{II} 2PCF predicted by our model is effectively undetectable with current or near-future surveys using galaxy background spectra. A robust detection of the IGM-only \ion{Mg}{II} 2PCF would require a new generation of spectroscopic surveys combining (i) substantially higher spectral resolution, (ii) per-pixel S/N far exceeding those achievable in $\lesssim 10\rm~h$ exposures on 4m-class telescopes, and (iii) a vastly larger number of bright background sightlines than planned for existing facilities. In practice, such a program is beyond the scope of the current JUST survey and would likely require dedicated more powerful, massively multiplexed, and/or very high resolution spectroscopy on larger apertures (e.g., the ELT/MOSAIC and ANDES, \citealt{Hammer2014,Marconi2022,DOdorico2024}).

On the other hand, our conclusions apply only to the diffuse IGM component (\textsection\,\ref{subsec:MockGasCube}). As shown by \citet{Tie2024}, the observed \ion{Mg}{II} 2PCF exhibits a much stronger small-scale peak when discrete CGM absorbers are present, with amplitudes of $\xi_{\rm peak}\sim10^{-3}$ that are readily detectable in current data and vanish once these strong absorbers are masked. Our modeling suggests that a similar CGM-dominated 2PCF signal should be observable with JUST. In addition, as demonstrated in \textsection\,\ref{subsec:MgIIstacking}, the diffuse IGM \ion{Mg}{II} absorption may still be accessed through \emph{direct stacking} as a function of galaxy or large-scale structure environment. Therefore, while the 2PCF method developed by \citet{Hennawi2021} in its original form is an elegant unbiased theoretical probe of IGM metal enrichment, our simulations indicate that, for the foreseeable future with 4m telescopes, it will be most powerful as a tool to characterize the \ion{Mg}{II} absorbers in the CGM instead of the IGM.

\section{Summary and Prospects} \label{sec:SummaryProspects}

In this work we have outlined and quantified the scientific potential of a coordinated Euclid/CSST/JPCam/JUST survey of EDF-N, aimed at probing the multi-phase CGM and IGM at the cosmic noon. We combined realistic source population models, mock photometry, and forward-modeled spectroscopy to assess, in a self-consistent framework, how well this joint survey can (i) characterize galaxy environments and large-scale structures, (ii) detect diffuse CGM emission, and (iii) carry out IGM tomography with background galaxy absorption lines.

On the photometric side, we adopted evolving galaxy and AGN LFs representative of $z\sim1-4$ and showed that the EDF-N contains a very large number of galaxies and a smaller but non-negligible population of AGN bright enough to be detected by Euclid, CSST, and JPCam, and to serve as potential background sources for spectroscopy with CSST and JUST. Down to a reference magnitude $m_{\rm ref}\simeq 23.5$ in an optical/near-IR band (observed frame), a single JUST FOV ($\sim1.4\rm~deg^2$) contains of order $\sim350$ background galaxies at $z\gtrsim1.5$, well below the $\sim2000$ available fibers while maintaining a high density of sightlines sampling the spatial distribution of the foreground gas structures. The Euclid VIS+NIR and CSST broad-band photometry already provide accurate photometric redshifts for galaxies at $z\sim1-2$, but our forward simulations demonstrate that the addition of the JPCam narrow bands around the Balmer $\sim4000~\text{\AA}$ break and the [\ion{O}{II}] doublet further increases the photo-$z$ accuracy to $\Delta v\sim10^2\rm~km~s^{-1}$ in a reasonable range of galaxy magnitudes and at $z\sim1.0-1.4$, especially for star-forming galaxies. This greatly enhances our ability to identify galaxy groups and (proto-)clusters, characterize the large-scale environment, and select clean samples of foreground and background galaxies tailored to CGM and IGM studies.

We then investigated the detectability of diffuse [\ion{O}{II}] emission from the CGM with JPCam narrow-band imaging. Assuming a simple model in which a fraction $f_{\rm CGM}\simeq0.3$ of the total [\ion{O}{II}] luminosity emerges in an exponential halo with scale length $r_{\rm s}\simeq30\rm~kpc$, we combined the JST250/JPCam sensitivity in the relevant filters with SFR distributions derived from our photo-$z$ experiments. For a single galaxy, even the most SF active systems ($\rm{SFR}\gtrsim100\rm~M_\odot~\rm{yr}^{-1}$) have extended [\ion{O}{II}] emission below the detection limit in a $\sim10\rm~h$ JST250/JPCam exposure. However, we showed that stacking moderate number of SF galaxies feasible to be selected from the EDF-N can yield statistically significant detections of the CGM out to $r\sim(50-100)\rm~kpc$.

On the spectroscopic side, we constructed a mock 3D \ion{H}{I} distribution in a fiducial redshift slab $1.0<z<1.4$, representing the IGM as a lognormal field that matches the mean Ly$\alpha$ optical depth and variance from hydrodynamic simulations. The comoving volume and spatial resolution of this cube were chosen to match the EDF-N area and the expected sightline density of background galaxies. Using bagpipes-based spectral models for a realistic background galaxy population, we generated mock CSST slitless spectra including Ly$\alpha$ absorption from the 3D gas cube and mock JUST fiber spectra including associated \ion{Mg}{II} doublet absorption. These synthetic spectra incorporate the instrumental line-spread functions, wavelength dependent throughputs, and noise properties appropriate for the planned deep CSST and JUST observations. The resulting catalog contains mock spectra for $\sim5.4\times10^3$ background galaxies at $z\geq1.5$. The 3D gas cube, the full galaxy catalog, and the background galaxy mock spectra catalog are all provided as supplementary data products for further analyses.

We applied the AOD method to the mock CSST Ly$\alpha$ forest spectra to reconstruct the underlying \ion{H}{I} distribution. By converting the normalized CSST spectra into apparent optical depths and integrating over redshift, we obtained 1D $N_{\rm HI}$ profiles along each LOS, which were then interpolated onto the 3D grid to map the gas distribution. Comparison of the reconstructed and true \ion{H}{I} fields shows that the AOD method recovers the main filaments and voids on scales comparable to the mean LOS separation, while small-scale features (e.g., the CGM) below the transverse sampling scale are naturally smoothed out. The global normalization is reasonably accurate, with a mean ratio $\langle N_{\rm HI,AOD}/N_{\rm HI,true}\rangle\simeq0.88$, and the expected trends are present: low-$N_{\rm HI}$ sightlines tend to be underestimated owing to noise and the non-linear mapping from flux to optical depth, while high-$N_{\rm HI}$ (partly saturated) sightlines are affected by unresolved substructure and peculiar velocities.

We also explored the prospects for measuring IGM metallicity with \ion{Mg}{II} absorption in the JUST spectra. First, we attempted to detect diffuse IGM \ion{Mg}{II} via direct stacking of normalized JUST spectra of absorbers. In an idealized experiment where the absorber positions are known perfectly, we find that the \ion{Mg}{II} doublet could be firmly detected. However, when absorbers are selected from noisy, low-resolution CSST Ly$\alpha$ forest spectra with realistic redshift uncertainties, the stacked \ion{Mg}{II} doublet signals are washed out. We therefore conclude that, at the planned depth, the combination of CSST and JUST is unlikely to detect diffuse IGM \ion{Mg}{II} absorption, unless the redshift of the absorbers can be determined accurately in other ways, although stronger CGM-dominated signals associated with foreground galaxies remain promising.

We further adapted the 2PCF method developed by \citet{Hennawi2021} to our mock JUST spectra, treating the \ion{Mg}{II} absorption as a continuous forest and using the peak at the doublet separation to infer the large-scale mean \ion{Mg}{II} opacity. For the diffuse IGM component alone, our simulations show that the expected 2PCF signal is extremely weak for the planned JUST (or with other 4m telescope) survey setup: even with $\sim5,000$ sightlines, $\sim10\rm~h$ exposures, and covering the full EDF-N area, the predicted S/N is far below what would be required for a robust detection. Simple scaling arguments indicate that neither modest increases in exposure time nor reasonable enlargements of the survey area can bridge this gap. A statistically significant detection of an IGM-only \ion{Mg}{II} 2PCF would require a new generation of spectroscopic surveys combining substantially higher spectral resolution, S/N, and/or larger number of bright background sightlines than planned for JUST.

Overall, our analysis demonstrates that the joint Euclid/CSST/JPCam/JUST programme in EDF-N will be highly effective in (i) mapping galaxy environments and large-scale structures at the cosmic noon, (ii) detecting and characterizing extended [\ion{O}{II}]-emitting CGM halos in stacked narrow-band imaging, and (iii) reconstructing the large-scale \ion{H}{I} distribution from low-resolution slitless CSST spectra using the AOD technique. By contrast, constraining the metallicity of the diffuse IGM through \ion{Mg}{II} absorption appears to be beyond the reach of the current survey configuration, although the same dataset will be powerful for an accurate measurement of the foreground galaxy redshift and studying CGM-dominated \ion{Mg}{II} signals.

Looking ahead, the methodology developed here can be readily extended to other photometric and spectroscopic surveys. Forthcoming wide-field optical and near-IR imaging surveys such as LSST, Roman, WFST, and extended J-PAS-like surveys will deliver deeper and more homogeneous photometry over much larger areas, enabling the identification of galaxy structures and background-source populations for IGM tomography over a broad range of redshifts. Combining such photometric data with next-generation massively multiplexed spectrographs on 4-8m-class telescopes (e.g., DESI-like facilities, Subaru/PFS, 4MOST, WEAVE, etc.) will make it possible to carry out similar tomographic experiments in multiple independent fields, to probe different environments, and to cross-correlate the cool/warm gas traced in Ly$\alpha$ and metal lines with hot gas seen in future X-ray surveys (e.g., with AXIS; \citealt{Koss2025}) and cold gas mapped in \ion{H}{I} 21~cm line with the SKA and its pathfinders (e.g., \citealt{Duffy2012}).

The ultimate gains, however, may come from the next-generation 39m European Extremely Large Telescope (E-ELT) equipped with both multi-object and high-resolution spectrographs such as MOSAIC \citep{Hammer2014} and ANDES \citep{Marconi2022,DOdorico2024}. With two orders of magnitude larger light collecting power, the ELT-class facilities will reach a much higher S/N for comparably bright background galaxies, at even higher spectral resolutions. This is not only powerful for a finer IGM tomography with the Ly$\alpha$ forest \citep{Japelj2019}, but may also allows us to detect metal absorption lines from the diffuse IGM. The improvement is not only based on the improved detection limit of metal lines, but also because of a better resolved Ly$\alpha$ forest (typically at higher redshifts of $z\gtrsim2.3$) so a more accurate location of the absorber (MOSAIC has a resolution of $R\sim5,000-20,000$). With a multi-object spectroscopic survey covering a moderate sky area (e.g., $\sim0.5\rm~deg^2$ with MOSAIC), either the direct stacking or the 2PCF of metal absorption line doublets (e.g., \ion{Mg}{II} or \ion{C}{IV}; \citealt{Tie2022,Tie2024}) should become powerful probes of the diffuse IGM metallicity. Furthermore, special case studies of individual bright background quasars with extremely high-S/N, high-resolution spectroscopy (e.g., ANDES has $R\simeq100,000$) will help us to directly detect the IGM absorbers and characterize their column density distributions (e.g., \citealt{Songaila2005,LiJ2025,Yu2025}). The Euclid/CSST/JPCam/JUST strategy developed in this paper thus provides both a concrete science case and a set of optimization tools for planning future multi-wavelength, multi-facility programmes that will ultimately link galaxy evolution to the assembly and enrichment of the IGM at the peak epoch of cosmic star formation.

\begin{acknowledgements}
The authors acknowledge the assistance of ChatGPT (OpenAI, San Francisco, USA) in coding and improving the clarity and grammar of the manuscript. All scientific content, analysis, and conclusions were conceived and verified by the authors.
J.T.L., X.H.Y., and X.D.Y. acknowledge the financial support from the science research grants from the China Manned Space Program with grant no. CMS-CSST-2025-A04. J.T.L. also acknowledge the support from the China Manned Space Program with grant no. CMS-CSST-2025-A10, the National Science Foundation of China (NSFC) through the grants 12321003 and 12273111, and Jiangsu Innovation and Entrepreneurship Talent Team Program through the grant JSSCTD202436. 
R.A.D. acknowledges support from the Conselho Nacional de Desenvolvimento Cient\'{i}fico e Tecnol\'{o}gico -CNPq through BP grant 312565/2022-4.
Y.G. acknowledges the support from the CAS Project for Young Scientists in Basic Research (No. YSBR-092), and the science research grant from the China Manned Space Project with grant No. CMS-CSST-2025-A02.
X.H.Y. also acknowledge the science research grants from the China Manned Space Project with no. CMS-CSST-2021-A02. This project is also supported in part by Office of Science and Technology, Shanghai Municipal Government (grant Nos. 24DX1400100, ZJ2023-ZD-001).
\end{acknowledgements}

\bibliographystyle{raa}
\bibliography{CSSTEuclidJUSTJPAS}

\appendix

\section{Creating the Mock 3D Gas Distribution Data Cube}\label{Appsec:3DGasCube}

We adopt a flat $\Lambda$CDM cosmology with the \citet{Komatsu2011} WMAP7 parameters. All distances quoted in this section are comoving unless otherwise noted.

\subsection{Gaussian random field and lognormal density}

Following the IGM-tomography methodology of \citet{Japelj2019}, we represent the large-scale gas distribution by a smoothed Gaussian random field with a controlled correlation length. We first create a white-noise field $\delta_{\rm white}(\chi, x, y)$ with zero mean and unit variance on the $N_\chi \times N_y \times N_x$ grid. We then apply an isotropic Gaussian filter in Fourier space:
\begin{equation}
  \tilde{\delta}_{\rm sm}(\mathbf{k}) = \tilde{\delta}_{\rm white}(\mathbf{k})
  \,\exp\!\left[-\frac{1}{2}k^2 L_{\rm corr}^2\right],
\end{equation}
where $\mathbf{k}=(k_\chi,k_y,k_z)$, $k^2 = k_\chi^2 + k_y^2 + k_z^2$, and $L_{\rm corr}$ is the comoving correlation length (we adopt $L_{\rm corr} \simeq 2.5$~Mpc, comparable to the effective smoothing lengths used in Ly$\alpha$ forest tomography reconstructions; e.g.\ \citealt{Japelj2019,Lee2018}). The smoothed field is then transformed back to real space,
\begin{equation}
  \delta_{\rm g}(\chi, x, y) = \mathcal{F}^{-1}\bigl[\tilde{\delta}_{\rm sm}(\mathbf{k})\bigr],
\end{equation}
and renormalized to have zero mean and unit variance.

To obtain a physically motivated, positive-definite gas density field and mimic the non-linear density distribution of the IGM, we map the Gaussian field to a lognormal density \citep[e.g.][]{Japelj2019}:
\begin{equation}
  \rho(\chi, x, y) = \exp\!\bigl[\delta_{\rm g}(\chi, x, y) - \tfrac{1}{2}\sigma_{\rm g}^2\bigr],
\end{equation}
where $\sigma_{\rm g}^2$ is the variance of $\delta_{\rm g}$. We then normalize $\rho$ so that its volume-averaged value satisfies $\langle\rho\rangle = 1$. The dimensionless overdensity is
\begin{equation}
  \delta(\chi, x, y) = \rho(\chi, x, y) - 1.
\end{equation}
This procedure produces a web-like structure whose filaments and sheets have characteristic comoving thickness $\sim L_{\rm corr}$ and which is adequate for studying line-of-sight absorption statistics and large-scale trends, even though it is not realistic and does not attempt to model smaller, gravitationally bound systems such as individual galaxies and (proto-)clusters.

\subsection{Ly$\alpha$ optical depth field}

We convert the overdensity field into an approximate Ly$\alpha$ line-center optical depth field $\tau(\chi, x, y)$ using a simple power-law relation adapted from the fluctuating Gunn-Peterson approximation:
\begin{equation}
  \tau(\chi, x, y) = \tau_0\,\bigl[1 + \delta(\chi, x, y)\bigr]^{\alpha_\tau},
\end{equation}
with parameters $(\tau_0,\alpha_\tau)$ chosen to yield realistic contrast in the Ly$\alpha$ forest-like structures. In the current implementation we adopt:
\begin{equation}
  \tau_0 = 0.4, \qquad \alpha_\tau = 1.6,
\end{equation}
which produce optical-depth fluctuations of order unity across the cosmic web, broadly comparable to the structures considered in Ly$\alpha$ tomography studies \citep[e.g.][]{Japelj2019,Lee2018}. The cube stores $\tau(\chi, x, y)$ as the Ly$\alpha$ line-center optical depth per voxel.

For visualization and later comparison, we also compute the dimensionless optical-depth contrast: 
\begin{equation}
  \Delta_{\tau}(\chi, x, y) \equiv \frac{\tau(\chi, x, y)}{\langle\tau\rangle} - 1,
\end{equation}
where $\langle\tau\rangle$ is the volume-averaged optical depth over the entire cube.

\subsection{Conversion to \ion{H}{I} column density}

We further convert the line-center optical depth field $\tau$ into a neutral hydrogen column density per voxel, $N_{\rm HI}(\chi,x,y)$, assuming that the Ly$\alpha$ absorption in each cell can be approximated by a Doppler-broadened line with a uniform $b$-parameter. For a transition with rest-frame wavelength $\lambda_0$ and oscillator strength $f$, the Ly$\alpha$ line-center optical depth is: 
\begin{equation}
  \tau_0 = N_{\rm HI}\,\sigma_0,
\end{equation}
where the line-center cross-section is:
\begin{equation}
  \sigma_0 = \frac{\sqrt{\pi} e^2}{m_e c}\,\frac{f\,\lambda_0}{b}.
\end{equation}
Here $e$ is the electron charge, $m_e$ the electron mass, $c$ the speed of light, and $b$ the Doppler parameter. We adopt Ly$\alpha$ parameters:
\begin{equation}
  \lambda_0 = 1215.67~{\rm \AA}, \qquad f = 0.4164,
\end{equation}
and a fiducial Doppler parameter $b=27~{\rm km~s^{-1}}$. Inverting the above relation yields
\begin{equation}
  N_{\rm HI}(\chi,x,y) = \tau(\chi,x,y)\,\frac{m_e c}{\sqrt{\pi} e^2}\,
  \frac{b}{f\,\lambda_0}.
\end{equation}

\begin{figure*}
  \centering
  \includegraphics[width=1.0\textwidth]{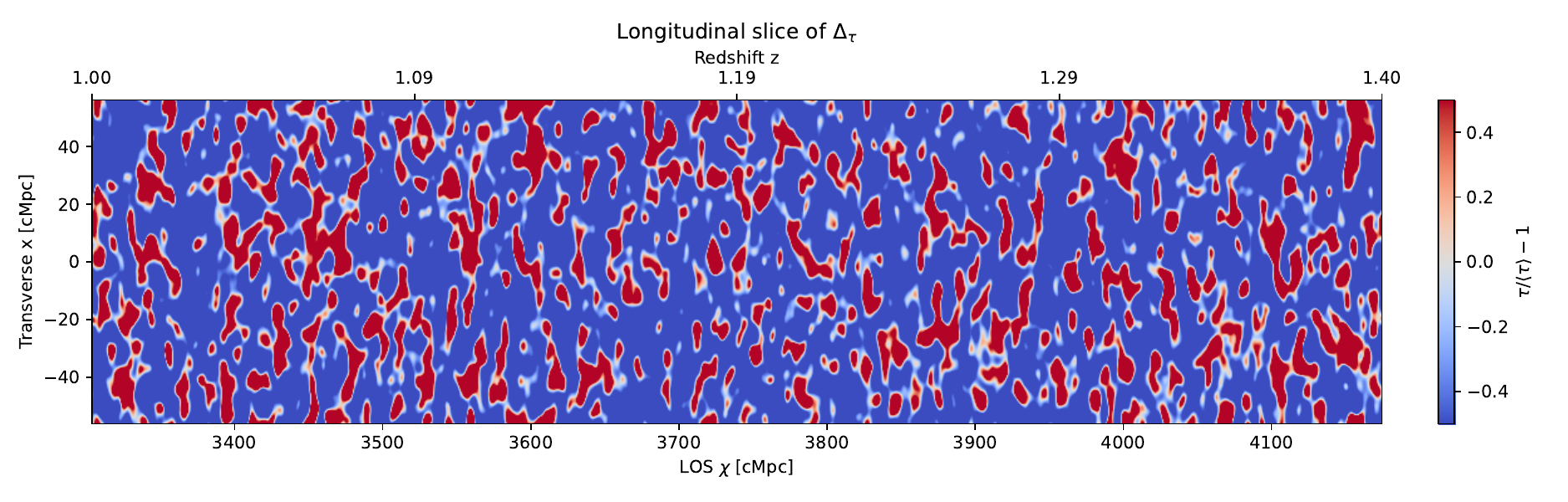}
  \caption{Diagnostic longitudinal slice in the $\chi$ (or $z$)--$x$ plane through the simulated Ly$\alpha$ optical depth contrast $\Delta_{\tau}$ (Eq.~\ref{eq:tauconstrast}) cube. This slice is extracted from the 3D $\Delta_{\tau}$ data cube by averaging over a finite band (10~cMpc) in the transverse $y$ direction. It is plotted as a function of redshift (horizontal axis) and transverse coordinate $x$ (vertical axis).}
  \label{fig:tausliceszx}
\end{figure*}

\begin{figure}
  \centering
  \includegraphics[width=0.49\textwidth]{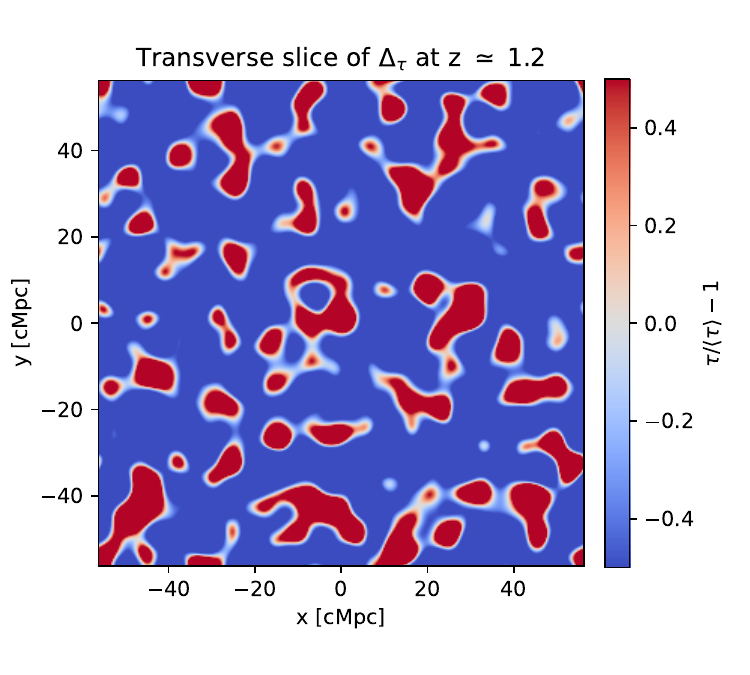}
  \caption{Similar as Fig.~\ref{fig:tausliceszx}, but showing a transverse slice of $\Delta_{\tau}$ at $z\simeq 1.2$ in the $x$--$y$ plane.}\label{fig:tauslicesxy}
\end{figure}

\subsection{FITS cube products}

We store both the Ly$\alpha$ optical depth and the derived column density in FITS cubes with minimal but useful World Coordinate System (WCS) information. The primary Header Data Units (HDUs) have data arrays of shape $(N_\chi, N_y, N_x)$ and store
\begin{itemize}
  \item $\tau(\chi,x,y)$ as the Ly$\alpha$ optical depth per voxel, in a file \\[1ex]
        \texttt{gas\_tau\_cube\_step1.fits}, with \texttt{BUNIT} set to \texttt{"optical depth"};
  \item $N_{\rm HI}(\chi,x,y)$ as the Ly$\alpha$ column density per voxel, in a file \\[1ex]
        \texttt{gas\_NHI\_Lya\_cube\_step1.fits}, with \texttt{BUNIT} set to \texttt{"cm-2"}.
\end{itemize}
In each file, a simple 3D WCS is defined with axes $(\alpha,\delta,z)$, assuming a tangent-plane projection centered at an arbitrary reference position $(\alpha_0,\delta_0)$. The angular pixel scales in right ascension and declination are derived from the comoving pixel sizes $(\Delta x,\Delta y)$ at $z_{\rm mid}\simeq1.2$, and the redshift axis is linearly spaced between $z_{\min}=1.0$ and $z_{\max}=1.4$. For convenience, a secondary binary table extension named \texttt{AXES} stores the exact 1D arrays of $z$, $\chi(z)$, $x$, and $y$ (all in comoving Mpc for the spatial coordinates), which can be used to reconstruct the mapping between voxel indices and physical positions.

To visually inspect the simulated large-scale structures and verify that the chosen parameters produce realistic filamentary patterns, we generate diagnostic 2D slices from the 3D cube. In each case, we use the contrast fields $\Delta_{\tau}$ and $\Delta_{\log N}$ rather than the raw $\tau$ or $\log N_{\rm HI}$, as described in \textsection\,\ref{subsec:MockGasCube}. The $\Delta_{\log N}$ maps are presented in the main text (Figs.~\ref{fig:NHIsliceszx}, \ref{fig:NHIslicesxy}), while the $\Delta_{\tau}$ maps are presented here in the appendix (Figs.~\ref{fig:tausliceszx}, \ref{fig:tauslicesxy}).

\section{Spectral Catalog for Background Galaxies at $z\geq1.5$}\label{Appsec:SpecCatBckGal}

We extend the single-sightline experiment discussed in \textsection\,\ref{subsec:CoupleGasGalaxy} to the entire background galaxy population and construct a homogeneous spectral catalog for all mock background galaxies (not including AGN) at $z_{\rm g} \ge 1.5$. The resulting data product, \texttt{step4\_5\_bg\_spectra\_cube.fits}, contains one row per galaxy and is designed to be the main input for subsequent statistical analyses (e.g., \textsection\,\ref{subsec:AODCSSTreconstruction}, \ref{subsec:MgIIstacking}, \ref{subsec:MgII2PCF}).

Starting from the background galaxy catalog constructed in \textsection\,\ref{subsec:MockGalaxyCat}, we select all galaxies with $z_{\rm g} \ge 1.5$ (tabbed as $z_{\rm s}$ in the catalog). For each selected background galaxy we read the stellar mass $M_\star$, SFR, and comoving position $(\chi_{\rm Mpc}, x_{\rm Mpc}, y_{\rm Mpc})$ from this galaxy catalog, then follow the method introduced in \textsection\,\ref{subsec:CoupleGasGalaxy} to generate mock CSST and JUST spectra by combining the background galaxy spectrum to the absorption lines produced by its foreground gas distribution ($z_{\rm gas}=1.0-1.4$).

The final product is an Astropy table in FITS format, with one row per background galaxy and the following main groups of columns:

\begin{itemize}
  \item \textbf{Parameters of the galaxies:} ID, redshift $z_{\rm s}$, stellar mass $\log M_\star$, SFR, SFR factor $f_{\rm SFR}$, reference magnitude $m_{\rm ref}$ (rest-frame UV band), CSST $z$-band magnitude $m_z$, and comoving position $(\chi_{\rm Mpc}, x_{\rm Mpc}, y_{\rm Mpc})$.
  
  \item \textbf{Integrated gas column densities:} total neutral hydrogen column density $N_{\rm HI,tot}^{1.0-1.4}$ and total \ion{Mg}{II} column $N_{\rm Mg\,II,tot}^{1.0-1.4}$ along the sightline in the redshift range $z=1.0-1.4$.

  \item \textbf{CSST spectra:} for each band (GU, GV, GI) and each galaxy we store the wavelength grid and three spectra: (i) the model spectrum without absorption, (ii) the model spectrum with Ly$\alpha$ absorption, and (iii) a noisy realization of the absorbed spectrum assuming a deep CSST survey depth ($8\times250\rm~s$ exposure).

  \item \textbf{JUST spectra:} similarly, for JUST we also store the wavelength grid and the three spectra (with subscript ``J''), but using \ion{Mg}{II} doublet absorption lines and 10~hr exposure instead.
\end{itemize}

This catalog provides a self-consistent set of mock spectra and associated physical and geometric parameters for all background galaxies, enabling both individual sightline studies and statistical analyses such as stacking Ly$\alpha$ and representative metal-line absorption as a function of galaxy properties and environment.

\section{3D \ion{H}{I} Data Cube Reconstructed from the Mock CSST Spectra with the AOD Method}\label{Appsec:AODGasCubeData}

We also make the AOD reconstructed 3D \ion{H}{I} data cube $N_{\rm HI}^{\rm AOD}(\chi,x,y)$ in \textsection\,\ref{subsec:AODCSSTreconstruction} available as a supplementary data product. The cube is stored as a single \texttt{FITS} file with one primary image extension containing a four–dimensional array whose axes correspond to absorber redshift $z$, comoving LOS distance $\chi$ (or equivalently the index along the redshift grid), and the two transverse comoving coordinates $(x,y)$ on a regular grid of $(2048, 512, 512)$ cells. The file header additionally records the 1D $z$ and $\chi$ grids and the transverse coordinates in physical units, together with the cosmological parameters used to compute $H(z)$ and $\chi(z)$. This data product allows users to repeat our validation tests, experiment with alternative transverse reconstructions (e.g., Gaussian smoothing instead of Voronoi assignment), and combine the AOD cube with other simulated tracers of the IGM.

\end{document}